\documentclass[useAMS,usedcolumn,usenatbib,usegraphicx]{mn2e}
\usepackage{times}

\usepackage{aas_macros}
\usepackage{textcmds}
\usepackage{multirow}
\usepackage{xcolor}
\usepackage{hyperref}
\usepackage{mathabx}
\usepackage{ctable}
\newcommand\T{\rule{0pt}{2.6ex}}       
\newcommand\B{\rule[-1.2ex]{0pt}{0pt}} 

\title[Methanol Along the Path from Envelope to Protoplanetary Disc]{Methanol Along the Path from Envelope to Protoplanetary Disc}

\author[Maria N. Drozdovskaya et al.]{Maria~N.~Drozdovskaya$^{1}$\thanks{E-mail: drozdovskaya@strw.leidenuniv.nl}, Catherine~Walsh$^{1}$, Ruud~Visser$^{2}$, Daniel~Harsono$^{1,3}$\newauthor
 and Ewine~F.~van~Dishoeck$^{1,4}$\\
$^{1}$~Leiden Observatory, P.O. Box 9513, 2300 RA, Leiden, The Netherlands\\
$^{2}$~Department of Astronomy, University of Michigan, 500 Church Street, Ann Arbor, MI 48109, U.S.A.\\
$^{3}$~SRON Netherlands Institute for Space Research, P.O. Box 800, 9700 AV Groningen, The Netherlands\\
$^{4}$~Max-Planck-Institut f\"{u}r Extraterrestrische Physik, Giessenbachstrasse 1, 85748 Garching, Germany.}

\begin{document}

\date{Accepted 2014 Month xx. Received 2014 Month xx; in original form 2014 Month xx}

\pagerange{\pageref{firstpage}--\pageref{lastpage}} \pubyear{2014}

\maketitle

\label{firstpage}

\begin{abstract}
Interstellar methanol is considered to be a parent species of larger, more complex organic molecules. A physicochemical simulation of infalling parcels of matter is performed for a low-mass star-forming system to trace the chemical evolution from cloud to disc. An axisymmetric 2D semi-analytic model generates the time-dependent density and velocity distributions, and full continuum radiative transfer is performed to calculate the dust temperature and the UV radiation field at each position as a function of time. A comprehensive gas-grain chemical network is employed to compute the chemical abundances along infall trajectories. Two physical scenarios are studied, one in which the dominant disc growth mechanism is viscous spreading, and another in which continuous infall of matter prevails. The results show that the infall path influences the abundance of methanol entering each type of disc, ranging from complete loss of methanol to an enhancement by a factor of $>1$ relative to the prestellar phase. Critical chemical processes and parameters for the methanol chemistry under different physical conditions are identified. The exact abundance and distribution of methanol is important for the budget of complex organic molecules in discs, which will be incorporated into forming planetary system objects such as protoplanets and comets. These simulations show that the comet-forming zone contains less methanol than in the precollapse phase, which is dominantly of prestellar origin, but also with additional layers built up in the envelope during infall. Such intriguing links will soon be tested by upcoming data from the \textit{Rosetta} mission.
\end{abstract}

\begin{keywords}
astrochemistry -- stars: protostars -- protoplanetary discs -- comets: general.
\end{keywords}

\section{Introduction}
\label{intro}

The birth of a star is accompanied by the emergence of a protoplanetary disc, which serves as a nursery for young planets. These systems are formed from dense, cold cores \citep{Shu1987}. With time, a rotating core evolves into the constituents of a protostar and a protoplanetary disc, to conserve angular momentum, leaving behind a remnant envelope. Young star-forming systems set the initial conditions for planet and comet formation, but it remains unclear what level of chemical complexity is attained in the early protoplanetary and cometary material. It is necessary to consider the chemical evolution as the envelope collapses in order to know the chemical composition of the regions in which protoplanets emerge. Complex organics formed at the early stages of star formation are likely important ingredients for the prebiotic chemistry of planetary systems.

Starting from simple chemical ingredients, such as water (H$_{2}$O) and carbon monoxide (CO), star-forming systems flourish in chemical complexity. Complex organic molecules have been observed in high-mass and low-mass protostars, as discussed in reviews by \citet{HerbstvD2009} and \citet{CaselliCeccarelli2012}. However, the mechanism for their formation remains a puzzle. Laboratory experiments of UV-irradiated methanol (CH$_{3}$OH) ice have shown that the radicals produced by the photodissociation of CH$_{3}$OH go on to produce species that are more complex, such as glycolaldehyde (CH$_{2}$OHCHO) and methyl formate (HCOOCH$_{3}$) (see, e.g., \citealt{Oberg2009}). Methanol is thus considered to be a vital precursor of complex organic molecules. Furthermore, laboratory studies demonstrate that CH$_{3}$OH can easily form under conditions as cold as $12$ K via grain-surface hydrogenation reactions \citep{WatanabeKouchi2002, Fuchs2009}, making it abundant ($\sim1-10$ per cent of water ice) under prestellar conditions \citep{Dartois1999, Gibb2004, Pontoppidan2004, Oberg2011c2d}. This suggests that methanol is readily available in prestellar environments to provide the key radicals to form larger, more complex organic molecules.

This paper explores the chemical history of the material that enters a protoplanetary disc by tracing the chemical evolution of infalling parcels of matter along trajectories from the early envelope into the disc. The aim is to gain insight into the evolving chemistry with changing physical conditions along infall paths. \citet{Visser2009}, henceforth referred to as V09, studied the chemical evolution of H$_{2}$O and CO ices in a similar manner. Their key conclusion was that large outer regions of the disc contain pristine water, i.e. H$_{2}$O that has never been sublimated, dissociated, reformed, nor refrozen on its path from cloud to disc. A follow-up study, \citet{Visser2011}, henceforth referred to as V11, discussed the full gas-phase chemistry. They concluded that comets form from material with different chemical histories. The physical model of V09 and V11 is a $2$D, semi-analytic simulation and is also the model used in this work. The motivation for $2$D physicochemical collapse models spawns from the successes of $2$D disc models (see, e.g., \citealt{AikawaHerbst1999}).

Over the years, a number of other codes have also been developed with their own advantages and disadvantages. $1$D physicochemical models (e.g., \citealt{Ceccarelli1996, Garrod2008}) are not able to treat the vertical structure of the disc realistically, while $3$D hydrodynamic simulations including radiative transfer and chemistry are computationally challenging. In fact, this has only been attempted recently up to the first hydrostatic core phase \citep{Furuya2012, Hincelin2013}, which is a transient pressure-supported stage with a lifetime of a few thousand years, after the onset of collapse, and prior to the formation of a true protostar \citep{Larson1969}. The results of this work are explicitly compared to the conclusions drawn by \citet{Hincelin2013} later in this paper, since both works analyse the survival of molecules with individual methods, but up to different stages of evolution.

In recent years, large data sets on ice observations in low-mass protostars have become available from dedicated \textit{Spitzer Space Telescope} and ground-based surveys. From these bigger samples, \citet{Oberg2011c2d} made an inventory of the major ice components and their variations from source to source. For the case of methanol, variations from $\lesssim1-25$ per cent with respect to H$_2$O ice are found, but the mean values and their spread are within a factor of two of those found in comets. Observational surveys have also been carried out for quiescent clouds prior to star formation \citep{Boogert2011, Boogert2013}, which have demonstrated that methanol ice is now commonly detected at the level of a few per cent with respect to water ice in cold clouds that have never been heated. Recently, a signpost of active methanol chemistry, the methoxy radical (CH$_{3}$O), was detected in a cold dark cloud \citep{Cernicharo2012}. Finally, an intriguing link between interstellar and cometary chemistry is suggested based on other data such as chemical complexity and observed isotope fractionations \citep{MummaCharnley2011}.

This work expands upon V11 by considering the evolution of more complex ices, using a significantly larger and more comprehensive gas-grain chemical network. The work presented here focuses on methanol to probe the budget of complex organic compounds for two different sets of initial physical conditions. The critical steps and processes in the methanol chemistry are also investigated.

The paper is structured as follows. The physical and chemical models and methods employed are described in Section~\ref{models}. The simulated physical and chemical results are shown in Section~\ref{results}. Three different infall trajectories per physical scenario are discussed in detail, and the methanol ice budgets and histories for both discs are investigated. The astrophysical implications of the obtained results are addressed in Section~\ref{implications}. Finally, the concluding remarks can be found in Section~\ref{conclusions}.

\section{Models}
\label{models}
\subsection{Physical framework}
\label{physframe}

\begin{figure}
 \centering
 \includegraphics[width=0.5\textwidth,height=\textheight,keepaspectratio]{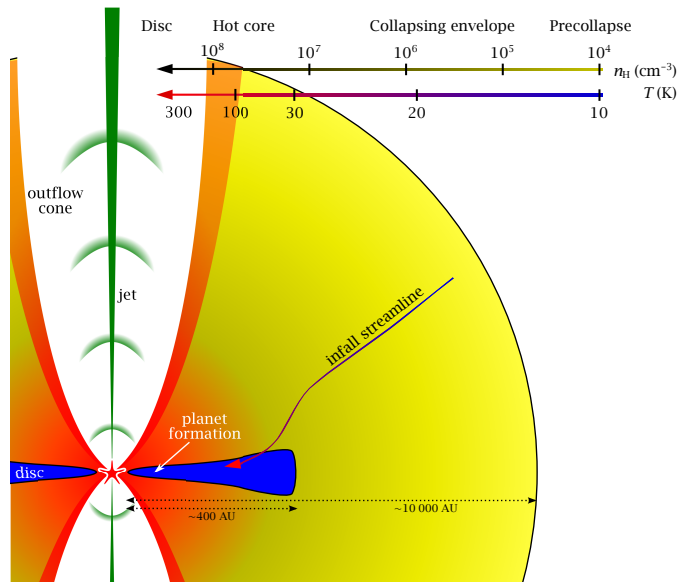}
 \caption{A cartoon depicting the physical components of an embedded phase of star formation, the planet-forming zone, and a trajectory of an infalling parcel. Typical gas densities, $n_{\rm H}$ (cm$^{-3}$), and the dust temperatures, $T_{\rm dust}$ (K), are also indicated. Figure not to scale (by R. Visser, adapted from \citealt{HerbstvD2009}).}
 \label{fgr:cartoon}
\end{figure}

\begin{figure}
 \centering
 \includegraphics[keepaspectratio]{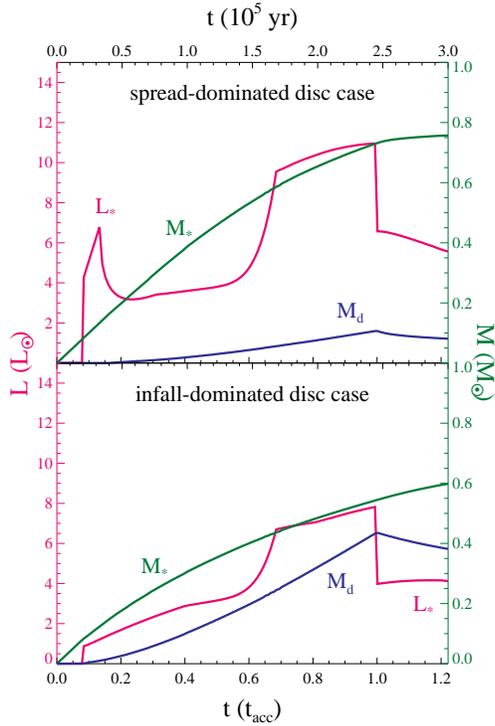}
 \caption{The time evolution of physical properties for the two cases studied. Each panel shows its respective evolution of stellar luminosity, $L_{*}$ ($L_{\Sun}$), stellar mass, $M_{*}$ ($M_{\Sun}$), and disc mass, $M_{d}$ ($M_{\Sun}$). The top panel displays the data for the spread-dominated disc case, i.e. case 3 from V09. The bottom panel corresponds to the infall-dominated disc case, i.e. case 7 from V09.}
 \label{fgr:starplot}
\end{figure}

\begin{figure}
 \centering
 \includegraphics[width=0.4\textwidth,height=\textheight,keepaspectratio]{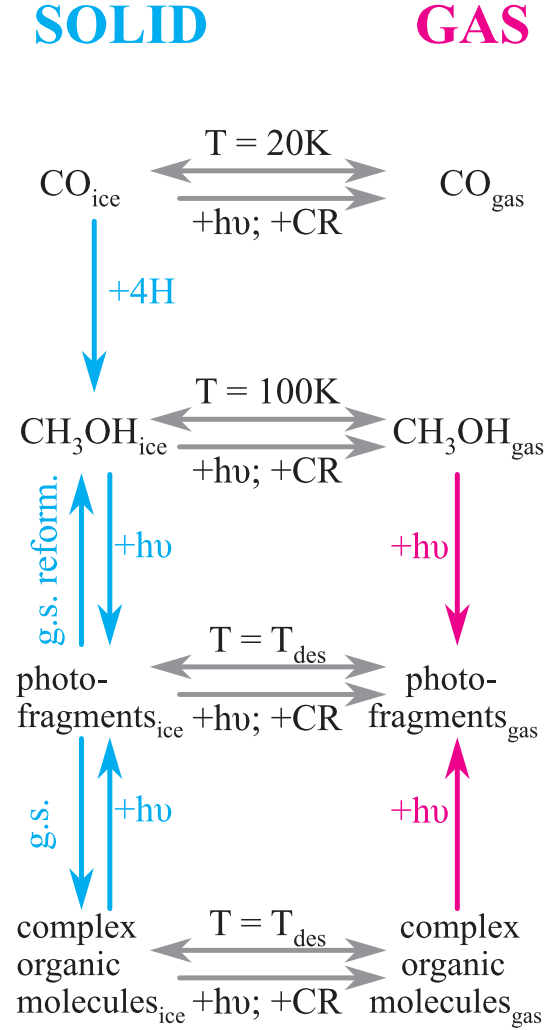}
 \caption{A schematic chemical network for methanol, summarizing the key reactions and processes at low temperatures ($T_{\rm dust}\lesssim100$ K). In the illustration, $h\nu$ corresponds to the energy of a photon, CR stands for a cosmic ray, g. s.  - for grain surface, g. s. reform. - for grain-surface reformation, and $T_{des}$ is the desorption temperature.}
 \label{fgr:network}
\end{figure}

\ctable[
 width = 0.5\textwidth,
 caption = Initial physical conditions\tmark,
 label = tbl:iniphys
 ]{@{\extracolsep{\fill}}llll}{
 \tnote{~$\Omega_{0}$: solid-body rotation rate; $c_{\rm s}$: effective sound speed; $M_{0}$: initial core mass; $t_{\rm acc}$: accretion time; $M_{\rm d}$: disc mass at $t_{\rm acc}$; $R_{\rm out}$: outer disc radius at $t_{\rm acc}$.}
 \tnote[b]{case 3 from V09}
 \tnote[c]{case 7 from V09}
 }{
 \hline
 Disc Case &  & spread-dominated\tmark[b] & infall-dominated\tmark[c] \T\B \\
 \hline
 $\Omega_{0}$ & $($s$^{-1})$ & $10^{-14}$ & $10^{-13}$ \T \\
 $c_{\rm s}$ & $($km s$^{-1})$ & $0.26$ & $0.26$ \\
 $M_{0}$ & $({\rm M}_{\Sun})$ & $1.0$ & $1.0$ \\
 $t_{\rm acc}$ & $($yr$)$ & $2.46\times10^{5}$ & $2.46\times10^{5}$ \\
 $M_{\rm d}$ & $({\rm M}_{\Sun})$ & $0.11$ & $0.44$ \\
 $R_{\rm out}$ & $($AU$)$ & $51$ & $294$ \B \\
 \hline}

\ctable[
 width = 0.4\textwidth,
 caption = The precollapse physical and chemical conditions$^{d}$,
 label = tbl:inichem
 ]{@{\extracolsep{\fill}}ll}{
 \tnote[d]{Initial abundances of F, Na, Mg, Si, P, S, Cl and Fe are all in the $10^{-8}-10^{-9}$ range relative to $n_{\rm H}$, see table $3$ of \citet{McElroy2013} and the references therein.}
 }{
 \hline
 $n_{\rm H}$ (cm$^{-3}$) & $4\times10^{4}$ \T \\
 $T_{\rm dust}$ (K) & $10$ \\
 $F_{\rm UV}$ (erg cm$^{-2}$ s$^{-1}$) & $0$ \\
 $n({\rm H})/n_{\rm H}$ & $5.0\times10^{-5}$ \\
 $n({\rm H}_{2})/n_{\rm H}$ & $5.0\times10^{-1}$ \\
 $n({\rm He})/n_{\rm H}$ & $9.8\times10^{-2}$ \\
 $n({\rm C})/n_{\rm H}$ & $1.4\times10^{-4}$ \\
 $n({\rm N})/n_{\rm H}$ & $7.5\times10^{-5}$ \\
 $n({\rm O})/n_{\rm H}$ & $3.2\times10^{-4}$ \B \\
 \hline}

\ctable[
 width = 0.5\textwidth,
 caption = {Select molecular abundances at the end of the precollapse phase of $3\times10^{5}$ yr and their binding energies},
 label = tbl:precollchem
 ]{@{\extracolsep{\fill}}lllr}{
 \tnote[e]{\citet{GarrodHerbst2006} - estimate}
 \tnote[f]{\citet{Fraser2001} - measurement for pure ice}
 \tnote[g]{\citet{BrownBolina2007} - measurement for pure ice}
 }{
 \hline
 Species & $n({\rm X_{gas}})/n_{\rm H}$ & $n({\rm X_{ice}})/n_{\rm H}$ & $E_{\rm des}({\rm X})$ (K) \T\B \\
 \hline
 H$_{2}$ & $5.0\times10^{-1}$ & $9.0\times10^{-7}$ &  $430^{e}$ \T \\
 CO & $3.7\times10^{-5}$ & $8.2\times10^{-5}$ & $1150^{e}$ \\
 H$_{2}$O & $5.1\times10^{-8}$ & $1.8\times10^{-4}$ & $5773^{f}$ \\
 H$_{2}$CO & $7.8\times10^{-9}$ & $6.9\times10^{-6}$ & $2050^{e}$ \\
 CH$_{3}$OH & $8.3\times10^{-11}$ & $8.0\times10^{-7}$ & $4930^{g}$ \B \\
 \hline
 }

The process of star formation is simulated using the model developed by V09 and V11, and further advanced by \citet{Visser2010} and \citet{Harsono2013}. This scheme is an axisymmetric $2$D, semi-analytic computation of the physical structure throughout the collapse and the disc formation stages. The collapse dynamics in the model are taken from \citet{Shu1977}. The effects of rotation are incorporated \citep{CassenMoosman1981, Terebey1984} and evolving outflow cavities are included. Magnetic fields are not considered.

The effects of any accretion shock at the disc surface are not taken into account, following the discussion in V09 (section 2.5). \citet{NeufeldHollenbach1994} investigated how interstellar dust is processed as it passes through an accretion shock. Vapourisation was determined to be the dominant mechanism of ice and grain destruction. They showed that the maximum grain temperature reached depends on the pre-shock velocities and densities. In the models presented here, the highest velocities and densities are reached at the earliest times, when the material accretes close to the protostar. At such positions, all ices would thermally desorb anyhow. Furthermore, this material is incorporated into the star before the end of collapse of the system. Sputtering of ices can also occur \citep{Jones1994}. The gas and dust that make up the disc experience a shock of at most $10$ km s$^{-1}$, which is not energetic enough to release strongly bound ices such as H$_2$O and CH$_3$OH. For the grain sizes assumed here, heating of the dust grains by the stellar UV photons dominantes over the accretion shock heating (V09). Thus, the release of weakly bound molecules, like CO, is still treated correctly in the formulation employed here.

With this model, $2$D time-dependent density and velocity distributions are obtained, with a self-consistent treatment of large-scale physical structures. The evolution of the central overdensity is simulated up to the formation of the first hydrostatic core for $2\times10^{4}$ yr and thereafter treated as a protostar (hereafter, the central overdensity throughout its evolution will be referred to as the star for simplicity). This allows for the determination of the stellar UV radiation field and thus the reaction rates of processes such as photodesorption, photodissociation, and photoionization. No external sources of radiation are included in this work since most star-forming regions are deeply embedded and thus shielded from external sources of UV photons. Accretion shocks onto the star, which are believed to be the source of excess UV around T Tauri stars \citep{Bertout1988}, are not considered. The central star thus solely controls the dust temperature, another important parameter for the chemistry. For further details, the reader should refer to the original publications on this model \citep{Visser2009, Visser2010, Visser2011, Harsono2013}.

This model is used to simulate the system up to the accretion time, $t_{\rm acc}$, defined as when the primary accretion phase onto the star ends and the outer shell of the envelope reaches the protoplanetary disc. This parameter is defined as:
\begin{equation}
 t_{\rm acc}=\frac{M_{0}G}{m_{0}c^{3}_{s}},
\end{equation}
where $M_{0}$ is the initial core mass, $G$ is Newton's gravitational constant, $m_{0}=0.975$ is a constant coming from the analytical solution of the hydrodynamics equations of a collapsing isothermal sphere \citep{Shu1977} and $c_{\rm s}$ is the effective sound speed.

The $3$D continuum radiative transfer code RADMC-3D\footnote[3]{\url{http://www.ita.uni-heidelberg.de/~dullemond/software/radmc-3d/}} is used to compute the dust temperature and the shielded stellar radiation in $2$D as a function of time based on the output of the collapse model. As a first-order approximation, it is assumed that the gas and dust are coupled, and thus the temperatures of both are equal. This assumption is most likely false for the outflow cavities, because the density is too low for gas-grain collisions to cool the gas efficiently, while the dust can still cool radiatively \citep{Draine1978, WeingartnerDraine2001}. The UV radiation calculated with RADMC-3D accounts for shielding and scattering by the material located between the star and the point of interest. A gas to dust mass ratio of $100$ is assumed. Opacity tables for icy grains from \citet{Crapsi2008} are used. In this work opacities for bare grains, dependent upon the dust temperature, are not incorporated. This is expected to cause temperature variations of at most $\sim10$ K around the $\sim100$ K zones, which is where the dominant ice component, H$_{2}$O ice, is sublimated for both the midplane and the surface layers of a disc (M.K. McClure, priv. comm.).

The collapse model computes trajectories of parcels of gas and dust as they fall in towards the star and into the protoplanetary disc, as illustrated in Fig.~\ref{fgr:cartoon}. Using the time-dependent velocity distribution, it is possible to follow material along infall streamlines. Each parcel's temporal and spatial coordinates can then be coupled to the corresponding values for physical parameters, such as density, dust temperature and stellar radiation, to which the parcels are exposed. With a sample of trajectories it is thus possible to trace the physical and chemical histories of various regions in the envelope and the protoplanetary disc. The motion of a large number of parcels from early to late times is shown in fig.~$7$ of V09.

The initial physical conditions control the subsequent evolution of the system. Here, two sets of initial parameters are considered, based on cases $3$ and $7$ from V09, respectively termed `spread-dominated' and `infall-dominated', and summarised in Table~\ref{tbl:iniphys}. Figure~\ref{fgr:starplot} shows select physical properties of the two scenarios. The two cases studied differ in the solid-body rotation rate, $\Omega_{0}$, by an order of magnitude. Since $t_{\rm acc}$ is independent of $\Omega_{0}$, $t_{\rm acc}$ is equal for the two cases under consideration in this work, at a value of $2.46\times10^{5}$ yr (Table~\ref{tbl:iniphys}). A slower solid-body rotation rate implies that more envelope material will be incorporated into the star and the disc will be less massive. Furthermore, since most of the material is used to build up the star, viscous spreading is the dominant mechanism of disc growth. As a result, the star formed in the spread-dominated disc case (case 3 from V09) is of higher mass and more luminous than that of the infall-dominated disc case (case 7 from V09), as shown in Fig.~\ref{fgr:starplot}.

On the other hand, the disc formed in the infall-dominated case is more massive than that formed in the spread-dominated case. The sizes of the discs vary by a factor of $\sim4$ at $t_{\rm acc}$ and the dominant motions that build the discs are not the same, which is reflected by the disc radial velocity profiles. This affects the trajectories of incoming parcels and the chemistry along them. For case $3$ from V09, which was also central to V11, the disc has an outer radius of $51$ AU and mainly grows by viscous spreading, thus termed `spread-dominated'. For case $7$ from V09 the disc has an outer radius of $294$ AU and primarily grows by the accretion of more envelope material onto it, thus termed `infall-dominated'. Due to the modifications between V09 and V11, the modelled protoplanetary disc sizes have decreased significantly (a factor $2-3$ difference in the outer disc radii for reasons explained in detail in \citealt{Visser2010} and \citealt{Harsono2013}). Due to the modifications in the model, it became necessary to modify the equation for the outflow cavity wall to:

\[ z=0.98\left(\frac{t}{t_{\rm acc}}\right)^{-3}R^{1.5}, \]

\noindent where $R$ and $z$ are cylindrical coordinates in AU. The full opening angle in our situation at $t_{\rm acc}$ is $11.58^{\circ}$ at $z=1000$ AU and $5.39^{\circ}$ at $z=10000$ AU, both of which are approximately a factor $3$ smaller than that in V11.

In this work the embedded phase of low-mass star formation is modelled, that is while the remnant envelope is still present. Few observations of such early discs are available, which makes it hard to constrain their dimensions. For example, the Keplerian disc in L1527 has been estimated to have an outer radius of $\sim125$ AU \citep{Tobin2012, Tobin2013, Sakai2014}. The set of Keplerian discs discussed in \citet{Harsono2014} have outer radii in the $\sim50-310$ AU range and masses varying from several thousandths to several tenths M$_{\Sun}$. By considering two different cases in this work with disc parameters in these ranges, future measurements are anticipated for confirmation.

\subsection{Chemical network}
\label{chemnet}

The physical model described above yields trajectories that trace streamlines of material infalling from the envelope into the protoplanetary disc. For simplicity, the chemical calculations are performed independently from the physical computations. The physical conditions (density, dust temperature, radiation field) at various time steps along the trajectories are used as input for the chemical code, which computes chemical abundances at each step. The procedure yields chemical abundances as a function of physical evolution for parcels probing various regions in the envelope and the protoplanetary disc, and is pictorially summarised in fig.~$1$ in V11.

The chemical model contains $666$ species and $8759$ reactions. The gas-phase network is the R\textsc{ate}12 release of the UMIST Database for Astrochemsitry (UDfA\footnote[5]{\url{http://www.udfa.net}}, \citealt{McElroy2013}). The network accounts for gas-phase two-body reactions. Three-body reactions are not considered, as they only become important at densities higher than $\sim10^{10}$ cm$^{-3}$, which are only attained briefly at the latest time steps in the evolution. Photoreactions (photoionization and photodissociation) by stellar and cosmic-ray-induced UV photons (generated by the cosmic-ray excitation of H$_{2}$, taken to be $10^{4}$ photons cm$^{-2}$ s$^{-1}$), and direct cosmic-ray ionization (with a rate of $5.0\times10^{-17}$ s$^{-1}$) for gas-phase species, are also included in R\textsc{ate}12. Photoreactions are computed according to equation 2 from V11, which takes the evolving stellar temperature into account. Self- and mutual shielding are taken into account for H$_{2}$, CO, and N$_{2}$ based on recent work \citep{Visser2009photodis, Li2013}. Grain-cation recombination is also included.

The model is supplemented with gas-grain interactions and grain-surface chemistry (and several additional reactions for complex organic molecules) extracted from the Ohio State University (OSU) network\footnote[6]{\url{http://www.physics.ohio-state.edu/~eric/research.html}} \citep{Garrod2008} and are calculated according to the detailed description in \citet{Walsh2014a} and the references therein. The chemistry is described by a two-phase model, i.e. gaseous and solid phases solely: that is, the ice surface and bulk are not treated as distinct phases. The rate equation approach is adopted for grain-surface reactions, based on \citet{HasegawaHerbstLeung1992} and \citet{HasegawaHerbst1993}. If one of the reactants is either an H or a He atom, then quantum tunnelling is allowed through the activation energy for a reaction and through diffusion barriers on the grains \citep{CazauxTielens2004, Watanabe2010}. For all other species, only classical hopping is permitted. The relation between the diffusion barrier ($E_{\rm diff}$) and the binding energy of a molecule to the surface ($E_{\rm des}$, which is also sometimes called the desorption energy) is taken to be $E_{\rm diff}=0.3\times E_{des}$. The set of binding energies compiled for use in conjunction with R\textsc{ate}12 is used with the exception of water ice. A higher value of $5773$ K from \citet{Fraser2001} for pure water ice is adopted instead.

The gas-grain interactions included are adsorption onto grain surfaces (also known as freeze-out) and thermal desorption \citep{HasegawaHerbstLeung1992, HasegawaHerbst1993}. Ices can also desorb non-thermally. Photodesorption (either by stellar or cosmic-ray-induced UV photons), cosmic-ray-induced thermal desorption (via heating of grains), and reactive desorption are all taken into account. The most recent experimental values for the photodesorption yields are adopted for the photodesorption rates \citep{Oberg2009bphotodes, Oberg2009aphotodes}. Furthermore, a coverage factor is used in light of recent experiments, which suggest that photodesorption occurs only from the top two monolayers \citep{Bertin2012}. The efficiency of reactive desorption is set to $1$ per cent \citep{Garrod2007, VasyuninHerbst2013}, but the efficiency of this process is not yet constrained by experiments and is likely variable, dependent on the reaction and the substrate (see e.g.~\citealt{Dulieu2013}).

Finally, grain-surface photoionization and photodissociation by stellar and cosmic-ray-induced UV photons are included. As a first order approximation, the equivalent rates for the gas phase are used. This is likely over-estimating the grain-surface photodissociation rates, since the mechanisms for UV photodissociation and photodesorption of ices are now understood to be related as demonstrated in molecular dynamics studies \citep{Andersson2006, AnderssonvD2008, Arasa2010, Arasa2011, Arasa2013, Koning2013} and experimental work \citep{Bertin2012, Fayolle2013}.

The grains are assumed to have a radius of $0.1$ $\mu$m and $n({\rm grains})/n_{\rm H}=2.2\times10^{-12}$, where by definition: $n_{\rm H}=n\left( {\rm H} \right)+2\times n\left( {\rm H_{2}} \right)$. The density of grain surface sites is $1.5\times10^{15}$ cm$^{-2}$ and the barrier thickness for quantum tunnelling between grain surface sites is taken to be $1$ ${\rm \AA}$, assuming a rectangular barrier \citep{HasegawaHerbstLeung1992}.

Prior to running the chemical model on a set of trajectories, it is necessary to obtain the initial chemical conditions at the onset of collapse. Assuming that the precollapse conditions are identical for the entire $2$D plane, the precollapse phase for a single point is simulated at constant physical conditions for $3\times10^{5}$ yr. The precollapse physical and chemical conditions are tabulated in Table~\ref{tbl:inichem}. The chemical abundances obtained at the end of the precollapse phase are used as initial chemical abundances for the trajectories. For reference, Table~\ref{tbl:precollchem} tabulates select molecular abundances at the end of the precollapse phase with their respective binding energies.

\subsection{Methanol chemistry}
\label{methchem}

The key chemical reactions involving CH$_{3}$OH are summarised in Fig.~\ref{fgr:network}. Early models suggested that methanol could form in the gas phase via ion-molecule reactions under dark cloud conditions \citep{MillarNejad1985, HerbstLeung1986}. This is thought to be a two-step process invoking initially radiative association \citep{Blake1987, Luca2002}, followed by dissociative recombination \citep{Geppert2006}. However, it was quickly suspected that this formation route is inefficient at low temperatures \citep{MillarLeungHerbst1987}. Later it was experimentally shown that, although fast, only $3$ per cent of the product channels of dissociative recombination lead to methanol \citep{Geppert2006}. Moreover, radiative association has to compete with other ion-molecule processes that have much larger rate coefficients resulting in too low gas-phase production of CH$_{3}$OH to explain dark cloud observations \citep{Garrod2006}.

Currently it is accepted that at low dust temperatures ($<100$ K and when external UV is negligible), grain-surface chemistry is responsible for the production of methanol via the sequential hydrogenation of CO:

\[ {\rm CO \stackrel{H}{\rightarrow} HCO \stackrel{H}{\rightarrow} H_{2}CO \stackrel{H}{\rightarrow} H_{3}CO/H_{2}COH \stackrel{H}{\rightarrow} CH_{3}OH,} \]

\noindent as first proposed by \citet{TielensHagen1982}. This mechanism has been extensively studied experimentally by various groups \citep{WatanabeKouchi2002, Watanabe2004, Hiraoka2002, Hidaka2004} and confirmed for $12$ K $\leq T_{\rm dust} \leq 20$ K \citep{Fuchs2009}. The second and fourth steps in the above mechanism are barrierless, because the H atoms are reacting with radicals. CO and H$_{2}$CO do not have unpaired valence electrons, thus the additions of H atoms are endothermic reactions. A reaction barrier of $E_{\rm A}=2500$ K for step one and for both routes (leading to either methoxy or hydromethoxy) of step three is adopted \citep{RuffleHerbst2001, Woon2002, GarrodHerbst2006}. This formation route is possible for dust temperatures as low as $10$ K, because quantum tunnelling allows the H atom to overcome this `large' barrier \citep{CuppenHerbst2007}. These processes are also reflected in the temperature-dependent `effective' reaction barrier measured in the laboratory, which are on the order of $\sim 400-500$ K \citep{Fuchs2009}, i.e. much lower than the theoretical value of $2500$ K. Above $\sim20$ K, the parent CO molecule sublimates from the grains and the residence time of the H atom on grain surfaces is too short for this reaction sequence to occur efficiently. As a result, methanol production significantly slows.

Similar to water and carbon monoxide, methanol can undergo thermal and non-thermal desorption, as shown in laboratory experiments. This work assumes $E_{\rm des}=4930$ K for CH$_{3}$OH, as stated in Table~\ref{tbl:precollchem}, which is the value measured for pure methanol ice \citep{BrownBolina2007}. At low temperatures ($T_{\rm dust}\lesssim100$ K) thermal desorption is inefficient for methanol, therefore under prestellar conditions it can only desorb non-thermally from the grains. Under dark conditions non-thermal desorption is sparked by the absorption of UV photons generated by the cosmic ray excitation of H$_{2}$ molecules. Weakly bound molecules like CO (see Table~\ref{tbl:precollchem}) are thought to be thermally desorbed due to the cosmic-ray heating of dust grains (at temperatures lower than their respective desorption temperatures). However, CH$_{3}$OH is a strongly bound species and its primary mechanism of coming off the grains is cosmic-ray-induced photodesorption \citep{HasegawaHerbstLeung1992, Shen2004, Roberts2007}.

Currently the chemical network includes the gas-phase and, in turn, the grain-surface photodissociation rates from R\textsc{ate}12. It is the photodissociation products of CH$_{3}$OH ice that are thought to pave the way to more complex species, such as glycolaldehyde and methyl formate, making this process and its parameters crucial for the chemistry. The photodissociation pathways of CH$_{3}$OH are:
\[ {\rm CH_{3}OH + h\nu \rightarrow CH_{3} + OH,} \]
\[ {\rm CH_{3}OH + h\nu \rightarrow CH_{2}OH + H,} \]
\[ {\rm CH_{3}OH + h\nu \rightarrow CH_{3}O + H,} \]
\noindent as investigated by \citet{GarrodHerbst2006}, \citet{Oberg2009}, and \citet{Laas2011}. These studies have analysed how the chemistry of methyl (CH$_{3}$), hydromethoxy (CH$_{2}$OH) and methoxy (CH$_{3}$O) radicals leads to complex species. In this work the `standard' branching ratios are adopted, namely $60:20:20$ per cent for the reactions in the order given above (table 1 in \citealt{Laas2011}). The exploration of the dependence of the formation of complex organics on the branching ratios in this model is a topic of upcoming work.

\subsection{Caveats}
\label{caveats}

Certain crucial chemical parameters affect the key chemical processes that govern the methanol abundance under different physical conditions. The first key aspect is the availability of reactant species (H, CO, HCO, H$_{2}$CO, CH$_{3}$O, CH$_{2}$OH) on the grains, which is determined by their binding energies. Methanol can only be formed if those species are present on the grains, and a change in the binding energies can influence the dominance of thermal and non-thermal desorption mechanisms. Secondly, these species need to be mobile in order to meet and react. The relation between $E_{\rm diff}$ and $E_{\rm des}$ is crucial here. In this work a factor of $0.3$ is used \citep{HasegawaHerbstLeung1992}, but other values also appear in the literature, e.g. $0.5$ in ~\citet{GarrodHerbst2006}. Hence, the mobility of individual species is related to the binding energy. Under cold conditions ($T_{\rm dust}\lesssim 20$ K), greater mobility could enhance the amount of methanol formed. However, under warmer conditions ($20~{\rm K}~\lesssim~T_{\rm dust}~\lesssim~100~{\rm K}$), the amount of methanol formed could be reduced if the photoproducts efficiently diffuse to form other species, rather than recombine immediately. This means that values of $E_{\rm des}$ for the photoproducts (CH$_{3}$, CH$_{3}$O, CH$_{2}$OH, OH) of methanol also need to be constrained. Thirdly, the efficiencies of the first and third hydrogenation steps en route to CH$3$OH, set by the value of $E_{\rm A}$, affect its production. The large difference between the theoretical and experimental (`effective') values has been pointed out in the previous section, and it is not clear how to treat this properly in models. Finally, the binding energy of methanol itself determines where it survives upon formation.

There are also two physical parameters that can have profound significance for the chemistry of methanol. The first one is the assumed dust model. Different values for the grain radius, the number density, the number of surface sites and the quantum barrier heights can affect adsorption rates of species onto the grains and the grain-surface diffusion rates \citep{HasegawaHerbstLeung1992}. Currently the dust grains are assumed to consist of compact spheres that are well-mixed with the gas. Aspects such as settling and dust coagulation have not been accounted for; however, they may be very important for adsorption and grain-surface associations due to the reduction of the grain surface area available for freeze-out and increased shielding from UV irradiation in the midplane \citep{AikawaNomura2006, Fogel2011, Vasyunin2011, Akimkin2013}. The other important physical parameter is the assumed ice model. Currently, frozen-out species are considered as a single entity, while it is known that ices are actually layered and that the mantle and the surface monolayers have different chemistry (see, e.g., \citealt{Ehrenfreund1998, Taquet2012}). For example, in clouds water freezes out first, followed by CO at later times. Methanol ice is thus expected to be sequestered and associated with CO rather than water ice \citep{Boogert2011}. Furthermore, accounting for the ice composition and structure would affect all of the chemical parameters mentioned above.

\section{Results}
\label{results}

In this work there are two discs with different dominant disc growth mechanisms, one that is spread-dominated and another that is infall-dominated. In Section~\ref{physevol} the physical evolution for these two cases is presented. In Section~\ref{chemevol} the chemical evolution is analysed. In each scenario, three trajectories entering the disc are studied in detail. Each disc has two trajectories with different methanol ice behaviours that are common to both cases, and one trajectory that is unique to its scenario. Finally, the methanol ice budget and history are presented for both cases at $\sim t_{acc}$ in Section~\ref{budgethistory}.

\subsection{Physical evolution}
\label{physevol}

\begin{table}
 \caption{Final parcel positions (at $t_{\rm acc}$)}
 \label{tbl:finpos}
 \begin{tabular*}{0.5\textwidth}{@{\extracolsep{\fill}}p{1cm}cp{3.8cm}rp{1cm}}
  \hline
  Case & Label & Parcel behaviour & $R$ $($AU$)$ & $z$ $($AU$)$ \T\B \\
  \hline
  \multirow{3}{*}{\begin{minipage}{1cm}spread-dominated disc\end{minipage}} & sA & devoid of CH$_{3}$OH$_{\rm ice}$ & 3.4 & 0.55 \T \\
    & sB & readsorbed prestellar CH$_{3}$OH$_{\rm ice}$ & 21.6 & 0.89 \\
    & sC & envelope-enriched CH$_{3}$OH$_{\rm ice}$ & 49.4 & 1.44 \B \\
  \hline
  \multirow{3}{*}{\begin{minipage}{1cm}infall-dominated disc\end{minipage}} & iA & devoid of CH$_{3}$OH$_{\rm ice}$ & 1.3 & 0.03 \T \\
    & iB & disc-formed CH$_{3}$OH$_{\rm ice}$ & 46.4 & 3.75 \\
    & iC & envelope-enriched CH$_{3}$OH$_{\rm ice}$ & 155.3 & 0.06 \B \\
  \hline
 \end{tabular*}
\end{table}

\begin{figure*}
 \centering
 \includegraphics[width=\textwidth,height=\textheight,keepaspectratio]{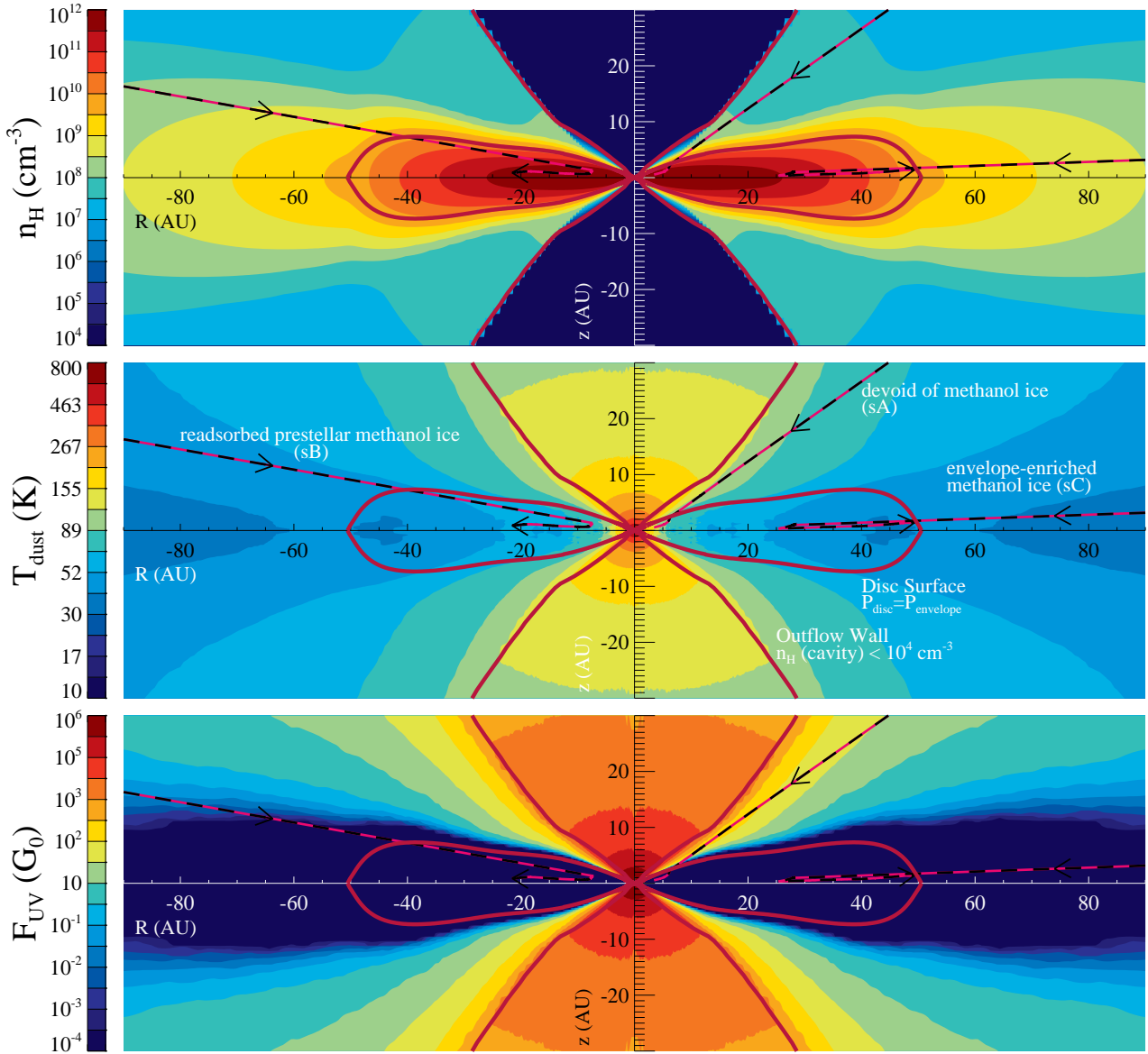}
 \caption{For the spread-dominated disc case from top to bottom: the gas density, $n_{\rm H}$ (cm$^{-3}$), the dust temperature, $T_{\rm dust}$ (K), and the combined field of the shielded stellar radiation and the cosmic-ray-induced UV field, $F_{\rm UV}$ ($G_{0}$). All these panels are at $\sim~t_{\rm acc}$ for the spread-dominated disc case. The disc surface and the outflow walls are labelled. Furthermore, three infall trajectories are depicted with black and pink lines as they move inwards from the far-out envelope at $t=0$ to the protoplanetary disc by $t_{\rm acc}$.}
 \label{fgr:physcase3}
\end{figure*}

\begin{figure*}
 \centering
 \includegraphics[width=\textwidth,height=\textheight,keepaspectratio]{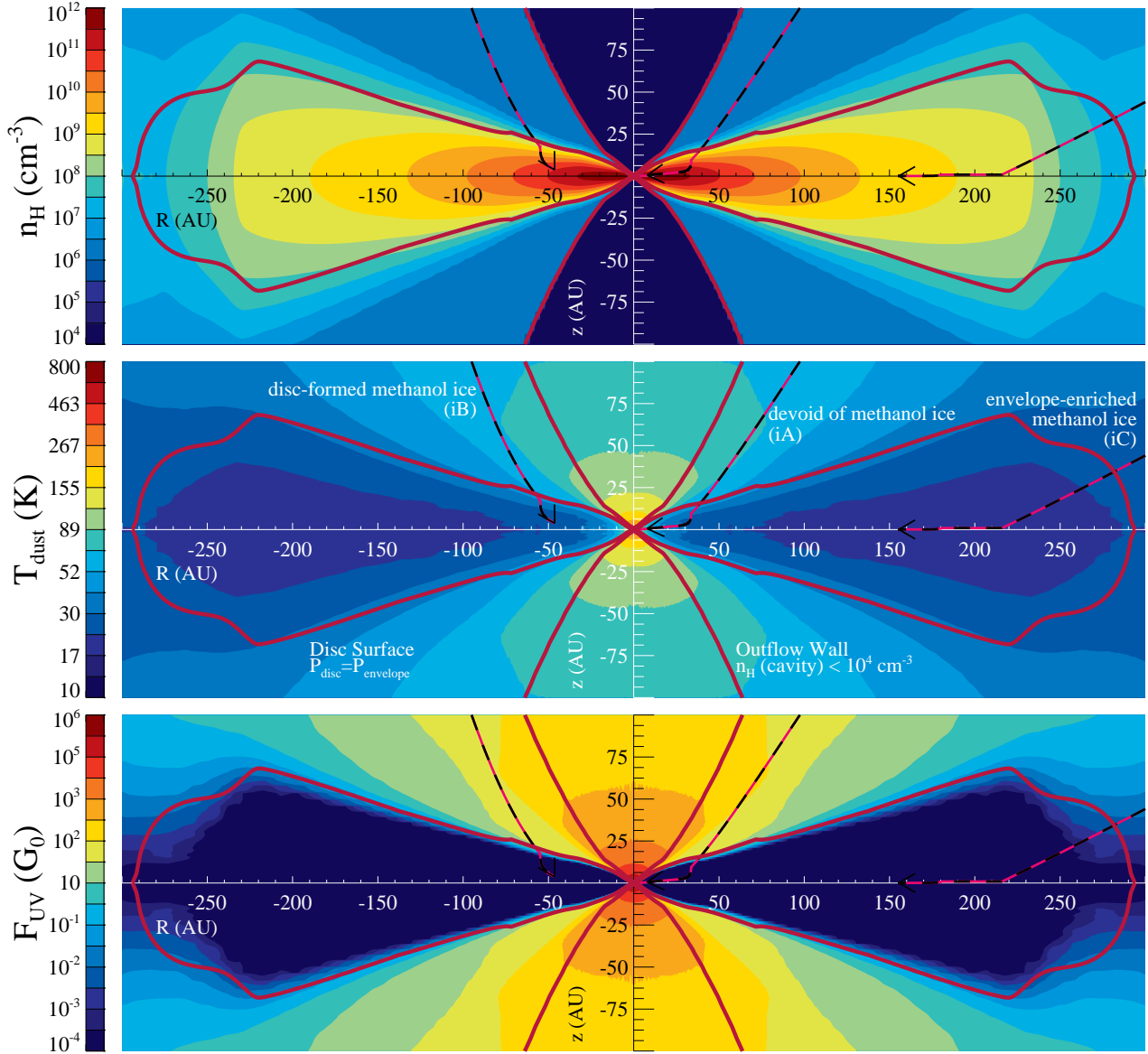}
 \caption{As Fig.~\ref{fgr:physcase3}, but for the infall-dominated disc case.}
 \label{fgr:physcase7}
\end{figure*}

\subsubsection{Spread-dominated disc case}
\label{physevolexp}

The top panel of Fig.~\ref{fgr:physcase3} shows the physical structure at the end of the simulation, at $\sim~t_{\rm acc}$, for the spread-dominated case. The star is at the origin and is surrounded by a protoplanetary disc with midplane densities of $\sim10^{12}$ cm$^{-3}$ at $8$ AU. The outflow cavities are identified as the regions where $n_{\rm H}<10^{4}$ cm$^{-3}$. The remnant envelope has densities in the $\sim10^{6}-10^{10}$ cm$^{-3}$ range. The disc surface is defined along the points where the envelope and disc pressures are equal, $P_{\rm envelope}=P_{\rm disc}$ \citep{Visser2010}.

The middle panel of Fig.~\ref{fgr:physcase3} shows the dust temperature at the same time step for the spread-dominated disc case, as computed with RADMC-3D (Section~\ref{physframe}). The highest temperatures are found within the outflow cavities, since this is where the density is lowest, and it is thus easier for the radiation to permeate and heat the dust. The central temperature (inner $0.1$ AU) exceeds $850$ K, while the midplane of the protoplanetary disc remains in the $\sim40-70$ K range depending on the radial distance from the star. Within $5$ AU, the dust temperatures in the disc exceed $100$ K. The envelope temperature varies in the $\sim20-110$ K range on larger scales and increases with proximity to the star. Due to the fact that the disc is only $51$ AU in radial size, it is easily heated passively by reprocessed stellar radiation and, consequently, is fairly warm (above $40$ K) in its entirety.

The bottom panel of Fig.~\ref{fgr:physcase3} shows the total strength of the far-UV $912-2066$ ${\rm \AA}$ ($6-13.6$ eV) flux, which has contributions from the shielded stellar radiation field and the cosmic-ray-induced UV field, at the same time step for the spread-dominated disc case. The strength of the interstellar UV radiation field, $G_{0}$, is $1.6\times10^{-3}$ erg cm$^{-2}$ s$^{-1}$ \citep{Habing1968}. The envelope regions closest to the outflow walls are subjected to a UV radiation field that is $1-100$ times stronger than the interstellar field and the outflow cavities are exposed to even stronger radiation ($>100$ $G_{0}$). The protoplanetary disc remains fully shielded from the stellar radiation ($A_{\rm V}\gtrsim10$ mag). Other regions are subject to weak UV irradiation, below the interstellar value. All zones of the system, even the midplane of the disc, are subject to a weak, but constant, cosmic-ray-induced UV radiation, which is assumed to be the typical $\sim 10^{-7}$ erg cm$^{-2}$ s$^{-1}$ \citep{PrasadTarafdar1983}.

All panels of Fig.~\ref{fgr:physcase3} depict three infall trajectories with black and pink lines as they move inwards from the far-out envelope at $t=0$ and into the protoplanetary disc by $t_{\rm acc}$. These three trajectories are representative of the key methanol ice behaviours encountered in the spread-dominated disc case and are labelled accordingly. This is further elaborated upon in Section~\ref{chemevol}. Table~\ref{tbl:finpos} lists the final positions of the three parcels at $t_{\rm acc}$, which are all within the protoplanetary disc. This disc grows primarily by viscous spreading, as reflected in the outward motion for two of the three trajectories after they enter the disc. A total of $250$ trajectories were calculated for the spread-dominated disc case, sampling the full spatial extent of the disc at $t_{\rm acc}$, and most of them embark on an outward path upon disc entry.

\subsubsection{Infall-dominated disc case}
\label{physevolinf}

The physical parameters for the infall-dominated disc case vary from those for the spread-dominated case presented previously. The top panel of  Fig.~\ref{fgr:physcase7} shows the density distribution at $\sim t_{\rm acc}$ for the infall-dominated disc case (on a much larger spatial scale than in Fig.~\ref{fgr:physcase3} due to the difference in disc sizes). The midplane densities of this much larger disc vary from $\sim10^{11}$ cm$^{-3}$ within the inner $30$ AU to $\sim10^{8}$ cm$^{-3}$ beyond $250$ AU. The outflow cavities are again low-density zones of $<10^{4}$ cm$^{-3}$. Note that the outflow cavities appear of different sizes between Fig.~\ref{fgr:physcase3} and Fig.~\ref{fgr:physcase7} only due to the difference in scales on the two figures. For this case the remnant envelope is of a lower density, $\sim 10^{5}$ cm$^{-3}$, which peaks at $\sim10^{8}$ cm$^{-3}$ close to the disc boundary.

The middle panel of Fig.~\ref{fgr:physcase7} displays the dust temperature at the same time step for the infall-dominated disc case. Overall this system is colder than the spread-dominated disc case, which correlates with this being a less massive star, and the disc being larger in size and mass. The disc has a large outer zone ($\gtrsim70$ AU) that is at $\sim20$ K. The temperature exceeds $100$ K only within $5$ AU, similar to the spread-dominated case, despite this disc having lower temperatures otherwise. The remnant envelope is generally cool and a dust temperature of $\sim 100$ K is exceeded only within $\sim30$ AU in $R$ and $z$. The outflow cavities are again the hottest regions with temperatures $> 150$ K close to the star ($z<23$ AU) and a central peak temperature (inner $0.1$ AU) of just under $800$ K.

The bottom panel of Fig.~\ref{fgr:physcase7} portrays the total strength of the far-UV flux at the same time step for the infall-dominated disc case. The outflow cavities are subject to the strongest radiation, $>1000$~$G_{0}$ within {$z\sim60$} AU. The disc remains shielded from the stellar UV flux and again subject only to the cosmic-ray-induced UV photon flux. However, in this case, a small disc surface layer exists, immediately below the labelled disc surface, that encounters UV flux at the interstellar level. Furthermore, larger regions of the envelope experience a UV flux $\sim 10$ $G_{0}$. Both effects are a consequence of the lower densities along the line of sight between the star and the remnant envelope (as seen in the top panel of Fig.~\ref{fgr:physcase7}). Additionally, this disc has a larger geometrical height due to its larger mass.

All panels of Fig.~\ref{fgr:physcase7} show three infall trajectories that are representative of the dominant methanol ice behaviours encountered for the infall-dominated disc case and are labelled as such. This is further discussed in Section~\ref{chemevol}. Table~\ref{tbl:finpos} lists the final positions of these parcels at $t_{\rm acc}$. This disc grows primarily by infall of matter as opposed to viscous spreading. Hence, none of the three trajectories have an outward component, and viscous spreading is observed only for a handful of parcels from the other 250 trajectories calculated for this case.

\subsection{Chemical evolution}
\label{chemevol}

\begin{figure*}
 \centering
 \includegraphics[width=\textwidth,height=\textheight,keepaspectratio]{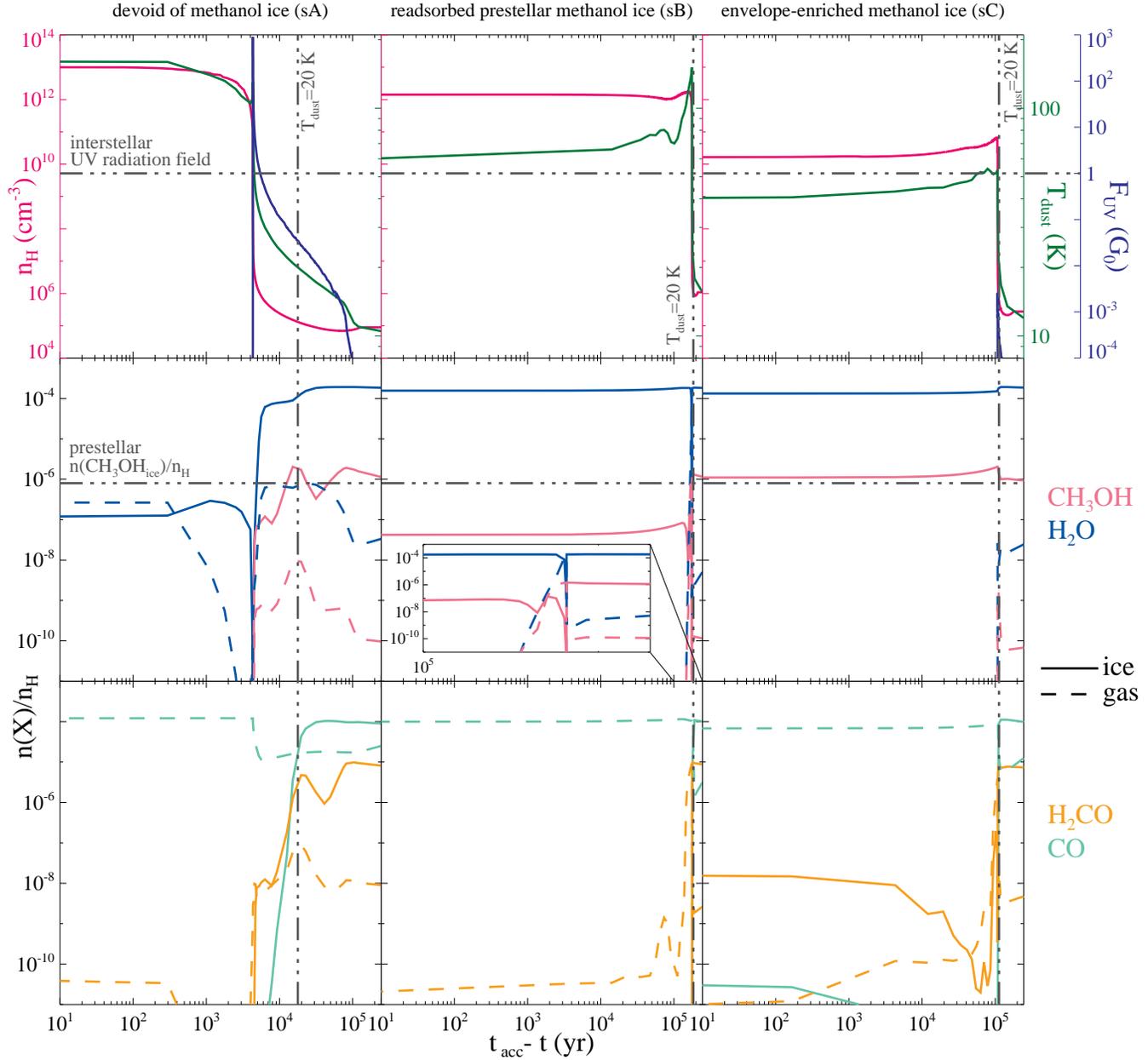}
 \caption{Physical conditions and molecular abundances as a function of time for the three parcels for the spread-dominated disc case. The figures should be read from right to left along the abscissa, which corresponds to going from early to late times. In the top three panels, the pink curves are the gas densities, $n_{\rm H}$ (cm$^{-3}$), the green curves are the dust temperatures, $T_{\rm dust}$ (K), and the dark blue curves are the combined fields of the shielded stellar radiation and the cosmic-ray-induced UV field, $F_{\rm UV}$ ($G_{0}$). The level of the interstellar UV radiation field is labelled, as are the points at which the $T_{\rm dust}=20$ K limit is surpassed, where CO ice desorbs. The lower six panels display the chemical abundances relative to $n_{\rm H}$ (see text). The colours correspond to different species, solid lines are used for the solid phase and dashed line for the gas phase. The initial prestellar methanol ice abundance is labelled as well. The center middle panel includes a blow-up of the critical transition region.}
 \label{fgr:chemcase3}
\end{figure*}

\begin{figure*}
 \centering
 \includegraphics[width=\textwidth,height=\textheight,keepaspectratio]{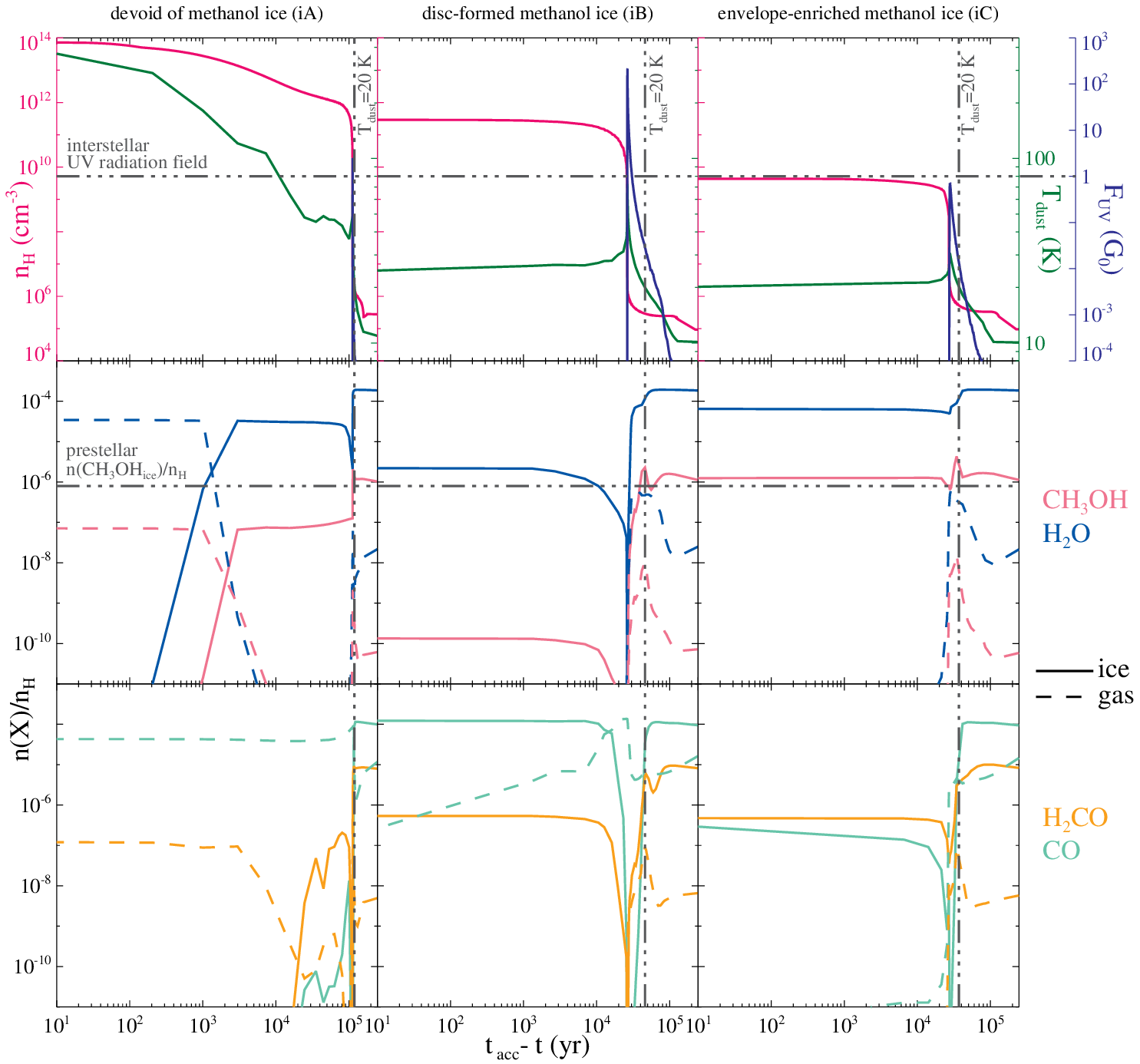}
 \caption{As Fig.~\ref{fgr:chemcase3}, but for the infall-dominated disc case.}
 \label{fgr:chemcase7}
\end{figure*}

\begin{figure*}
 \centering
 \includegraphics[width=\textwidth,height=\textheight,keepaspectratio]{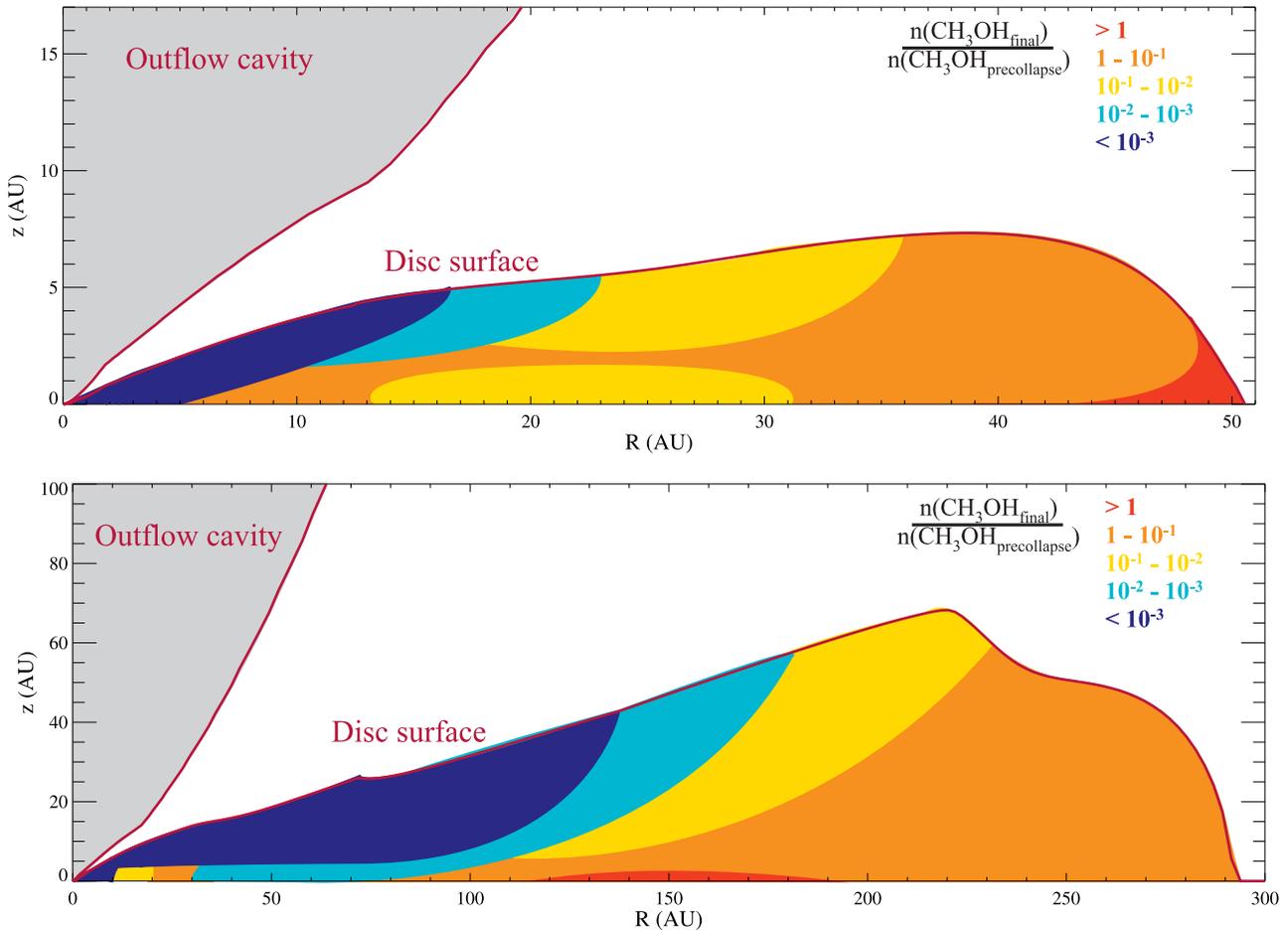}
 \caption{Methanol ice budget schematic for the spread-dominated disc case in the top panel and for the infall-dominated disc case in the bottom panel. The colours represent the value of the ratio of the methanol ice abundance at $\sim~t_{\rm acc}$ to the value at the onset of collapse (end of prestellar core phase) in the zones. The outflow cavity and the disc surface are also labelled. The zonal divisions are based on $\sim250$ trajectories per case.}
 \label{fgr:methbudget}
\end{figure*}

\begin{figure*}
 \centering
 \includegraphics[width=\textwidth,height=\textheight,keepaspectratio]{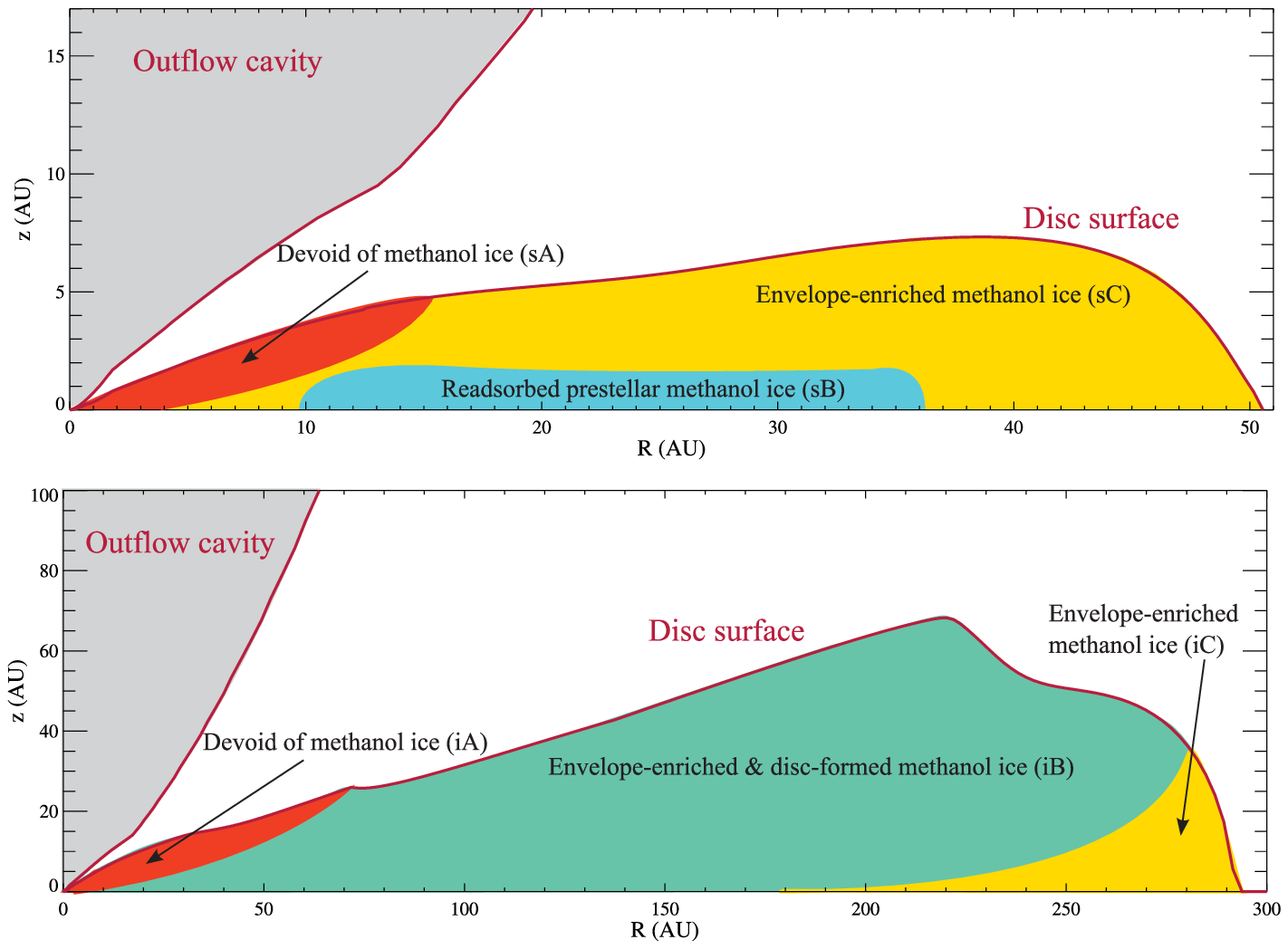}
 \caption{Methanol ice history schematic for the spread-dominated disc case in the top panel and for the infall-dominated disc case in the bottom panel. The zones are coloured depending on the characteristic methanol ice profiles for parcels building up those zones (see text) and are labelled accordingly. The zonal divisions are based on $\sim250$ trajectories per case.}
 \label{fgr:methhistory}
\end{figure*}

During the precollapse phase, CO forms via ion-molecule chemistry, reaching a peak canonical gas-phase abundance of $\sim10^{-4}$ relative to $n_{\rm H}$. (All abundances discussed hereafter are relative to $n_{\rm H}$ unless stated otherwise.) CO subsequently adsorbs onto the grains on a time-scale of $\sim10^{5}$ yr. Upon CO freeze-out, CH$_{3}$OH is efficiently formed on the grains from CO via sequential hydrogenation, as discussed in Section~\ref{methchem} \citep{TielensHagen1982}. This persists for $3\times10^{5}$ yr, the duration of the precollapse phase in the simulation. At the onset of collapse, the modelled methanol ice abundance is $8.0\times10^{-7}$, corresponding to $0.44$ per cent of water ice, which is consistent with observations of quiescent clouds without star formation \citep{Boogert2013}. Photodesorption by cosmic-ray-induced UV photons strips some of the ice from the dust grains, yielding a gas-phase methanol abundance of $8.3\times10^{-11}$, which is in agreement with $\sim10^{-10}$ from \citet{GarrodHerbst2006} and on the lower end of observed values $\sim10^{-8}-10^{-9}$ \citep{Gomez2011, Friberg1988}. Although, extraction of gas-phase abundances from observations remains challenging and needs to be approached cautiously.

The precollapse abundances obtained are used as the initial conditions for the computation of the chemical evolution along infall trajectories. Figure~\ref{fgr:chemcase3} shows select molecular abundances for the three parcels introduced in the previous sections for the spread-dominated disc case and Fig.~\ref{fgr:chemcase7} shows the same for the infall-dominated disc case. The figures should be read from right to left along the abscissa, which corresponds to decreasing time steps from early to late times. All six parcels start out with the same precollapse methanol abundance. As they undergo infall, the physical evolution they experience varies, which includes radically different temperatures, FUV fluxes and densities. These physical parameters are crucial for the chemistry and, as a result, the parcels obtain unique molecular abundance profiles. The figures should be analysed with the key reactions and processes, as depicted in the network in Fig.~\ref{fgr:network}, in mind. The time-scales of various processes are also of importance. For example, thermal desorption is very fast (on the order of $10^{3}$ yr for $n_{\rm H} \sim 10^{12}$ cm$^{-3}$, $T_{\rm dust} \sim 150$ K), while methanol ice grain-surface formation is slow (on the order of $10^{5}$ yr for $n_{\rm H} \sim 10^{4}$ cm$^{-3}$, $T_{\rm dust} \sim 10$ K). Besides methanol, Figs.~\ref{fgr:chemcase3} and \ref{fgr:chemcase7} shows the abundance of CO, H$_{2}$O and formaldehyde (H$_{2}$CO) for reference. Next, the chemical evolution is discussed for each characteristic methanol ice behaviour encountered.

\subsubsection{Enrichment of methanol ice}
\label{enrichment}

For the spread-dominated disc case the parcel with envelope-enriched methanol ice is depicted in the right column of Fig.~\ref{fgr:chemcase3} and labelled sC. It encounters densities in the $2.1\times10^{5}-7.1\times10^{10}$ cm$^{-3}$ range along its infall path with a steep jump of several orders of magnitude around $10^{5}$ yr before the end of the simulation, which corresponds to the parcel's entry into the disc. The temperature varies from $10$ K to its maximum value of $54$ K, followed by a decrease to $40$ K. The UV radiation encountered by the parcel briefly spikes up to a maximum of $2.6\times10^{-3}$ $G_{0}$. The temperature and the UV flux peak around the density peak, because immediately prior to disc entry is also where the parcel is closest to the star sans the shielding effects of the disc. Similar spikes in the physical parameters occur for all of the parcels considered; however at different times depending on when they enter the disc. After the disc entry, parcel sC remains inside the warm and heavily UV shielded disc, while moving outwards as the disc spreads viscously (Fig.~\ref{fgr:physcase3}).

Prior to the temperature reaching $20$ K, methanol ice continues to be formed efficiently on the grains via sequential hydrogenation of CO, which explains the initial rise in the methanol ice abundance. This occurs while the parcel is still infalling through the envelope, therefore the methanol in parcel sC is envelope-enriched. Once the temperature surpasses $20$ K, CO thermally desorbs off the grains and the formation of methanol slows down. Soon thereafter the temperature increases further, however it remains well below $100$ K, so methanol does not thermally desorb along this trajectory. The UV flux is also briefly elevated, however since its peak value is approximately four orders of magnitude lower than the interstellar value, neither photodesorption nor photodissociation remove significant amounts of methanol ice from the grains. 

Upon disc entry, the rapid and large density increase to $\sim10^{10}$ cm$^{-3}$ leads to the adsorption of gaseous methanol, which is reflected by a bump in the solid methanol abundance profile. The bump is levelled within $\sim10^{4}$ yr by the cosmic-ray-induced UV photons. Thereafter, the methanol ice abundance is preserved at the envelope-enriched level along the parcel's path through the disc. This is because the disc zones that the parcel encounters are all at temperatures below the desorption temperature of methanol and above the desorption temperature of CO. Furthermore, little stellar UV penetrates the disc. As a result, inefficient non-thermal destruction by the cosmic-ray-induced UV photons balances the slow formation of methanol ice via the OH and CH$_{3}$ route (whose abundance is very low due to the absence of stellar UV photons), the photodissociation (by the cosmic-ray-induced UV photons) of more complex species into CH$_{3}$OH and adsorption.

For the infall-dominated disc case the parcel with envelope-enriched methanol ice is depicted in the right column of Fig.~\ref{fgr:chemcase7} and is labelled iC. The density for this parcel is in the $9.5\times10^{4}-4.3\times10^{9}$ cm$^{-3}$ range. The temperature increases from $10$ K to a peak value of $31$ K, then decreases to $20$ K. The UV flux peaks at a value of $0.7$ $G_{0}$. This parcel encounters the lowest densities and temperatures, as it enters the outermost disc region considered (see, e.g., Fig.~\ref{fgr:physcase3} and Fig.~\ref{fgr:physcase7}).

The methanol ice profile of this parcel is very similar to that of parcel sC. In fact, the same processes are at play here. That is also the reason for this parcel being the envelope-enriched methanol ice analog for the infall-dominated disc case, although the parcel motion is very different. However, there is one key difference between parcels iC and sC, namely that for parcel iC the temperature drops below the desorption temperature of CO towards its final position. Once CO is again frozen out on the grains, methanol formation can occur. This is not seen here, because the simulation only runs for approximately another $10^{4}$ yr upon the readsorption of CO, which is not long enough for the grain-surface chemistry to produce a significant amount of methanol. In other words, parcels enter the outer disc later in the evolution and do not have enough time to form methanol within the disc.

\subsubsection{Destruction of methanol ice}
\label{destruction}

The left column of Fig.~\ref{fgr:chemcase3} depicts the parcel that is devoid of methanol ice for the spread-dominated disc case and labelled sA. It undergoes an increase in density in the $6.9\times10^{4}-1.0\times10^{13}$ cm$^{-3}$ range with a spike that coincides with a brief strong burst of UV radiation with a maximum value of $901$ $G_{0}$. The temperature again peaks prior to disc entry, but keeps rising to reach its maximum value of $161$ K at its final position. This is explained by the parcel's motion in the disc, which is inward in its entirety for all times, which, in turn, is a result of this parcel entering the disc within $10$ AU.

While the temperature is below $20$ K, more methanol is formed as was the case for parcels sC and iC with envelope-enriched methanol ice. However this parcel approaches the star much closer than the previous parcels and the temperature prior to disc entry peaks at $130$ K. Methanol ice is thereby rapidly removed by thermal desorption. Furthermore, the UV flux exceeds the interstellar value causing, not only photodesorption of methanol ice, but also photodissociation of both solid and gaseous methanol. This explains why the gas-phase methanol is also destroyed prior to disc entry. Once in the disc, the temperatures remain too high for any methanol to form. This parcel is devoid of not only methanol ice, but also of methanol gas.

The left column of Fig.~\ref{fgr:chemcase7} illustrates a parcel that is devoid of methanol ice for the infall-dominated disc case and labelled iA. The densities are in the $2.3\times10^{5}-7.7\times10^{13}$ cm$^{-3}$ range. The temperature, although containing certain spikes, tends to increase from $10$ K to higher values with the final point attaining the maximum of $444$ K. The UV spikes with a peak value of $2.5$ $G_{0}$.

This parcel is similar to parcel sA in the sense that they both have no solid methanol left at their final positions. However for parcel sA, thermal desorption, photodesorption and photodissociation ensured the destruction of methanol ice. Gaseous methanol was also photodissociated at the same time. In parcel iA, methanol ice undergoes pure thermal desorption, since the disc fully shields from stellar UV. As a result, gaseous methanol survives in the inner disc at the abundance level close to that of prestellar methanol ice.

\subsubsection{Readsorption of prestellar methanol ice}
\label{readsorption}

The parcel containing readsorbed prestellar methanol ice is depicted in the middle column of Fig.~\ref{fgr:chemcase3} and labelled sB. It is unique to the spread-dominated disc case. The density curve lies in the $8.4\times10^{5}-1.8\times10^{12}$ cm$^{-3}$ range. The temperature similarly rapidly spikes from $10$ to $151$ K, then gradually decreases to $59$ K. This parcel enters the disc earlier than parcel sC and also far away from the star ($R>25$ AU for both). This ensures that the UV flux remains at the cosmic-ray-induced level at all times. Contrary to the path of parcel sC, which initiates outward motion upon entry into the disc, this parcel first continues to move inwards, encountering high temperatures within the inner disc ($\lesssim10$ AU), while staying fully shielded from stellar UV, to only later embark on an outward journey.

Parcel sB spends little time below $20$ K due to its early disc entry. As a result, CO does not stay on the grains sufficiently long for any significant amount of additional methanol to be formed. The prestellar methanol is carried straight into the disc. However, once the parcel enters disc zones with temperatures higher than $100$ K, rapid thermal desorption of methanol occurs and all the ice is released into the gas. Once the parcel migrates outwards and the temperature drops again, methanol rapidly adsorbs onto the grains due to the high densities. The methanol molecules are still of prestellar origins hereafter; however, the abundance is lower than the initial prestellar value. This is due to the fact that some methanol was destroyed while it was in the gas-phase in the hot inner zones of the disc. The readsorbed prestellar methanol ice is preserved towards the final position of this parcel for the same reasons as it was preserved in parcel sC.

\subsubsection{Disc-formation of methanol ice}
\label{discformation}

The middle column of Fig.~\ref{fgr:chemcase7} depicts the parcel that contains disc-formed methanol ice and labelled iB. It is unique to the infall-dominated disc case. The densities lie in the $9.3\times10^{4}-2.9\times10^{11}$ cm$^{-3}$ range. The temperatures rise from $10$ K to a peak value of $73$ K, then decrease down to $24$ K. The UV flux briefly spikes at $215$ $G_{0}$.

Initially the methanol profile is similar to that seen for parcels sC and iC that contain envelope-enriched methanol ice, however before the disc entry the UV flux is $2$ orders of magnitude higher for this parcel. As a result, methanol gas is rapidly photodissociated. Methanol ice undergoes photodesorption, but most importantly also more frequent photodissociation. At the point of this strong UV spike, the temperature rises to $73$ K, therefore once a methanol ice molecule dissociates, it is much more likely that the photofragments thermally desorb off the grains rather than recombine. This leads to the rapid destruction of methanol ice, which is seen in the figure as a deep dip in the solid methanol abundance. The exact depth of the dip depends on the combination of the UV flux and the temperature. Once the parcel enters the disc, it is fully shielded from the stellar UV photons and the temperature decreases. In fact, the temperature drops to the level of $\sim20$ K and CO once again freezes out onto the grains. This in turn allows methanol formation via sequential hydrogenation of CO, but now within the disc. Due to the short time of $\sim10^{4}$ yr, the increase in methanol ice seen in the figure is slow and at a low abundance. The methanol ice found at the final position of this parcel is, in fact, partially disc-formed methanol ice.

\subsection{Methanol ice in discs}
\label{budgethistory}

\subsubsection{Methanol ice budget}
\label{methbudget}

To sample the two discs studied, around $250$ trajectories per case were computed. From this information approximate zones can be delineated in the discs based on the amount of methanol ice that they contain. This is done by means of schematics in the top panel for the spread-dominated disc case and in the bottom panel for the infall-dominated disc case in Fig.~\ref{fgr:methbudget}. The zones are coloured depending on the value of the ratio of the methanol ice abundance at $\sim~t_{\rm acc}$ to the constant value at the onset of collapse (end of the prestellar core phase, Table~\ref{tbl:precollchem}).

For the spread-dominated disc case in the top panel of Fig.~\ref{fgr:methbudget}, the inner $\sim5$ AU (in the midplane) is devoid of methanol ice, which is around the methanol snowline. Further out, there is radially dependent layering in the upper parts of the disc. The further away from the star, the less harsh the conditions (in terms of temperature and UV radiation) that parcels encounter upon infall and therefore, the more methanol ice is preserved on the grains. The outer zones of the disc lose the least methanol ice compared with that injected at the onset of collapse. (This holds for the precollapse abundances computed here.) The striking feature of this figure is the midplane, which contains less methanol ice than the prestellar core. This demonstrates the effect of the physics of the system on its chemistry. The trajectory undertaken by the parcels to get to the midplane in that radial range is by first approaching the star, and then viscously spreading outwards with the growing disc. For example, for a final position of $\sim20$ AU in the midplane, the approach must be as far in as $\sim8$ AU. Therefore, the parcels making up the midplane lose methanol ice during their approach by rapid thermal desorption. Once under methanol-ice preserving conditions again, methanol is quickly readsorbed onto the grains; however, at abundances lower than the initial prestellar value. Since the disc formed in this spread-dominated scenario is warm (above $\sim40$ K) methanol is not formed in the disc itself via the CO hydrogenation route. The CH$_{3}$ and OH route is viable at this temperature; however, these transient species are not efficiently formed due to the strong shielding from stellar radiation by the disc. The methanol ice formed via this route, the photodissociation (by the cosmic-ray-induced UV photons) of more complex species into CH$_{3}$OH and adsorption are balanced out by the inefficient non-thermal destruction of CH$_{3}$OH by the cosmic-ray-induced UV photons. It is the most outer disc zones that are rich in methanol ice due to cold, shielded conditions and further production in the envelope.

Comparing to the infall-dominated disc case in the bottom panel of Fig.~\ref{fgr:methbudget}, similarities and differences are present. Due to the lower remnant envelope densities in this case and therefore stronger UV flux at larger radii close to the disc surface, a much larger zone is methanol-ice poor. The methanol snowline still lies around $\sim5$ AU at the midplane. However, in this case the disc surface layers as far out as $\sim100$ AU lack solid methanol due to the strong FUV radiation, and thus due to rapid photodissociation, encountered immediately prior to entering the disc. Otherwise, a similar radially dependent layering still perseveres in the disc. The large outer zones of this infall-dominated disc contain more methanol ice than initially present in the system, as is the case for the spread-dominated disc for the same reasons. Contrary to the spread-dominated disc case, in this case, the disc is cold enough for CO to freeze out and methanol ice formation to initiate within the disc. Additionally, no methanol-ice poor midplane is seen. Here the midplane is populated by infalling parcels that are still methanol-ice rich, rather than those that have suffered methanol ice loss close to the star. Within the inner $\sim50$ AU the picture is more complicated, since different types of parcels flow inwards into that area. There are those from the surface layers that are methanol-ice poor, but there are also those from the outer zones, which are methanol-ice rich. 

\subsubsection{Methanol ice history}
\label{methhistory}

From the sets of trajectories computed, it is not only possible to compute the methanol budget, but also to deduce the methanol history in the disc. The select parcels discussed in Section~\ref{chemevol}, namely in Figs.~\ref{fgr:chemcase3} and~\ref{fgr:chemcase7}, display methanol ice profiles characteristic of various zones of the two discs. By classifying the profiles from all the parcels computed, the history of zones in the discs can be understood. The coloured regions are dominated by trajectories with the respective characteristic methanol ice profile, however this division is approximate.

The top panel of Fig.~\ref{fgr:methhistory} shows the methanol ice history for the spread-dominated disc case. The inner $\sim5$ AU midplane zone, containing parcels of type sA, is devoid of methanol ice, as anticipated from Fig.~\ref{fgr:methbudget}. The dominant portion of this disc contains envelope-enriched methanol ice, which is built up from parcels of type sC. Although all the parcels of this zone undergo methanol ice formation during infall, they do not necessarily contain more than the prestellar amount. Prior to entering the disc, some of them still lose methanol ice due to the temperatures and UV fluxes encountered, which is reflected in the budget schematic in Fig.~\ref{fgr:methbudget}. The third and final zone of this disc is the one containing readsorbed prestellar methanol ice and parcels of type sB. This zone corresponds to the methanol-ice poor midplane seen in Fig.~\ref{fgr:methbudget}.

In comparison to the conclusions drawn in V11 for water in their figs. $4$ and $6$, for the spread-dominated case, methanol is also absent within a similar inner disc zone. Furthermore, the readsorbed prestellar methanol ice in the midplane corresponds with the same behaviour seen for water by Visser et al. for a comparible region. The unique aspect of methanol ice is however the enrichment in the envelope en route to the disc, thanks to the formation pathway via CO. There are no analogous low-temperature formation pathways for water, and thus it is simply preserved from the prestellar phase into the outer disc zones, as seen in V11.

The bottom panel of Fig.~\ref{fgr:methhistory} depicts the methanol ice history for the infall-dominated disc case. Similar to the spread-dominated case and once again anticipated from the respective methanol budget figure, the inner $\sim5$ AU comprise the zone that is devoid of methanol ice and that contains parcels of type iA. The largest zone in this disc is that containing a mixture of disc-formed and envelope-enriched methanol ice, which is built up from parcels of type iB. This zone exists uniquely in this colder disc. The disc-formation and envelope-enhancement of methanol ice does not imply that there is more than the prestellar value. The methanol budget is not the same across this zone, as seen in Fig.~\ref{fgr:methbudget}. The last zone of this disc is associated with envelope-enriched methanol ice, which is where parcels of type iC come to reside.

\section{Astrophysical implications}
\label{implications}

Figures~\ref{fgr:chemcase3}-\ref{fgr:methhistory} show that the different physical conditions encountered along the various infall trajectories affect the chemical composition of the infalling material and set its history. The dust temperature and the UV radiation, in particular, drive the chemical changes. These two physical parameters determine the feasibility of critical chemical processes. Large variations in the density affect all processes and may result in, for example, rapid freeze-out. As a result, the material that enters a protoplanetary disc shows strong chemical differentiation according to regions. The inner $\sim4$ AU zone is expected to be methanol-ice poor in both discs studied and the prestellar fingerprint completely erased. Gas-phase preservation of the fingerprint is case-dependent. The extended outer regions, on the contrary, are methanol-ice rich and enriched during infall. In the case of the colder, infall-dominated disc, further enrichment occurs thanks to methanol ice formation within the disc itself. In the spread-dominated disc case, the midplane is methanol-ice poor, but does contain readsorbed prestellar methanol ice. In the infall-dominated disc case, the midplane is methanol-ice rich and contains both envelope-enriched and disc-formed methanol ice, depending on the time spent by each parcel in the disc.

In this work, regions have been identified where methanol is abundant. Methanol is thought to be a key precursor to larger, more complex organic molecules \citep{Oberg2009, Laas2011, Walsh2014a}. This means that one of the main ingredients is readily available in the extended outer regions of discs and in the midplane in the infall-dominated disc case. The formation and distribution of complex organic molecules during disc formation and envelope dissipation will be investigated in future work.

Several limitations and uncertainties in the chemical model were identified in Section~\ref{caveats}. The qualitative results presented are robust against chemical uncertainties, which are expected to have quantitative effects only. It is certain that the inner disc is methanol-ice poor, while the outer regions are methanol-ice rich. However, to what exact radial ranges the two belong is not definite. In addition, the physical model does not account for all known physical parameters, e.g., magnetic fields, viscous heating, and mixing are not included. It is also important to realise that the physical model is most useful for understanding what type of material is delivered to the early discs, and thus primarily serves for obtaining the initial conditions for other disc models that include more complete disc physics.

\subsection{Comparisons to previous works}
\label{comparisons}

In comparison with the 3D work of \citet{Hincelin2013}, our model runs up to $t_{\rm acc}=2.46\times10^{5}$ yr, while their simulations stop at the first hydrostatic core stage, namely at $\sim3.8\times10^{4}$ yr. As can be seen from Figs.~\ref{fgr:chemcase3} and~\ref{fgr:chemcase7}, prior to $(t_{\rm acc}-38\times10^{3}$~yr$)=2.08\times10^{5}$~yr profound chemical changes that occur thereafter are not probed. \citeauthor{Hincelin2013} concluded that the chemical composition in the outer disc is pristine while the temperature in the disc remains below the respective desorption temperatures. In this work, the outer parts of each disc have also been shown to preserve the prestellar methanol ice, but it is also further enriched en route through the envelope. Furthermore, it is found that photodesorption and photodissociation, in addition to thermal desorption, cause the destruction of methanol ice in the inner region of the spread-dominated disc. The UV field is also strong enough to destroy gas-phase methanol in that case. Thermal desorption is responsible for stripping methanol in the inner zone of the infall-dominated disc (since these parcels drift inwards along the shielded midplane rather than infalling into the disc under irradiated conditions). In the simulations of \citeauthor{Hincelin2013}, regions of the disc attain temperatures higher than the desorption temperature of methanol already as early as the first hydrostatic core. This is not the case in this work. Here the desorption temperature of methanol is only exceeded within the inner $\sim5$ AU at $t_{\rm acc}$. The reason for this is most likely the radiative transfer method. \citeauthor{Hincelin2013} make use of the flux-limited diffusion approximation, while here full continuum radiative transfer is performed.

The 2D hydrodynamical work of \citet{Brinch2008} obtained density and temperature profiles similar to those obtained in this work; however, their model did not include outflow cavities and was primarily focused on the envelope rather than the disc. The duration of their simulation is equal to that used here, namely $2.5\times10^{5}$ yr. The code of \citet{YorkeBodenheimer1999} adopted by \citeauthor{Brinch2008} used flux-limited diffusion, which results in a temperature difference of a few Kelvin in the envelope. The differences in disc temperatures, in comparison to the full continuum RADMC-3D calculation, cannot be quantified. \citeauthor{Brinch2008} also included accretion shocks onto the disc, which results in elevated temperatures ($\sim60$ K) along the disc surface. Their model was coupled with chemistry by \citet{vanWeeren2009}, in a similar approach to that performed here.

Van Weeren et al. used the R\textsc{ate}06 release of the UMIST database as their gas-phase network, i.e. the predecessor of the gas-phase network used here. For the surface reactions, \citeauthor{vanWeeren2009} used the network assembled in \citet{HasegawaHerbstLeung1992} and \citet{HasegawaHerbst1993} in conjunction with the modified reaction rates approach. Here, the larger, more up-to-date OSU network \citet{Garrod2008} is employed in combination with the classical rate equation method. The majority of the results presented in \citeauthor{vanWeeren2009} pertain to gaseous methanol, which becomes readily abundant at later times in the inner $\sim100$ AU at an abundance of $\sim10^{-5}$ due to thermal desorption. The only zone rich in gas-phase methanol seen in this work is the inner disc ($\sim5$ AU) in the infall-dominated disc case. This could be due to \citeauthor{vanWeeren2009} assuming $A_{V}=15$ for all times and all positions, thereby completely excluding photodesorption and photodissociation. In this work, these processes are carefully accounted for and are shown to play a crucial role. Alternatively, this could be due to their temperature structure in the disk, which must be much hotter than here in order to thermally desorb methanol from $\sim100$ AU inwards in the midpplane. According to table $2$ of \citeauthor{vanWeeren2009}, the amount of methanol ice decreases with time, but this work shows that this is only true for certain zones of the disc. Finally, their fig.~$19$ shows that methanol is not formed at any radial position in their simulation, which has, on the contrary, been seen for an array of positions in this work.

It was mentioned earlier that the results of this work are most useful in understanding the nature of the material entering the disc, rather than its composition as it evolves. As a result, the final abundances obtained are suitable as initial conditions for other more detailed disc models (e.g., \citealt{Walsh2014a}). In comparison to that work, which also looked at gas-phase and solid-state methanol among many other more complex species, there is agreement within an order of magnitude for the methanol ice abundance $\sim10^{-6}$. The main differences occur for the surface layers of the disc. In our models, the disc is still surrounded by remnant envelope material, which makes the surface layers less distinct, in particular for the spread-dominated disc case. In the work of \citeauthor{Walsh2014a} thermal and non-thermal processes result in a drop of around two orders of magnitude in the methanol ice abundance between the midplane and the disc surface (for $R\sim100$ AU). As for the midplane, \citeauthor{Walsh2014a} demonstrate the survival of methanol ice up to $\sim2$ AU, which then crosses over into a gas-phase methanol rich zone between $1$ and $2$ AU. This is similar to what is seen in this work for the infall-dominated case, however the methanol snowline lies around $\sim5$ AU here, which is the set by the adopted star-disc parameters. This is on the other hand contrary to what is seen with the spread-dominated case, in which gas-phase methanol is photodissociated once it comes off the grains. The stellar properties of \citeauthor{Walsh2014a} are that of a more mature system than the ones studied here, namely a $0.5$~$M_{\Sun}$ T Tauri star with an accretion rate of $10^{-8}$ $M_{\Sun}$ yr$^{-1}$.

\citet{Nomura2009} modelled the chemistry along accretion flows in the inner disc around a low-mass protostar. For the stream line shown in their work, which runs through the disc ($z=1.2H$, where $H$ is the disc scale height), a gas-phase methanol-rich zone is seen between $\sim5.5$ and $\sim6.5$ AU for an accretion rate of $1.0\times10^{-8}$ $M_{\Sun}$ yr$^{-1}$. \citeauthor{Nomura2009} concluded that for all species the existence of such gas-rich zones depends on the assumed accretion rate. In their models, gas-phase methanol is destroyed by reactions with ionised species or atomic hydrogen on time-scales of $\sim10^{4}$ yr. Thus, the size of the zone rich in gaseous methanol depends on whether the accretion flow is faster than this destruction time. For example, the gas-phase methanol-rich zone extends between $\sim2$ and $\sim6.5$ AU for a higher accretion rate of $5.0\times10^{-8}$ $M_{\Sun}$ yr$^{-1}$ (fig.~$3$ in \citealt{Nomura2009}). This is roughly in agreement with the infall-dominated disc case in this work, although their star is more massive and less luminous ($1.5$~$M_{\Sun}$ and $\sim5$~$L_{\Sun}$ vs. Fig.~\ref{fgr:starplot}). Here, the accretion rate is time-dependent and decreases from $\sim4\times10^{-6}$ $M_{\Sun}$ yr$^{-1}$ at the beginning of the simulation to $\sim2\times10^{-6}$ $M_{\Sun}$ yr$^{-1}$ at $\sim t_{\rm acc}$ for the infall-dominated case. This is around $2$ orders of magnitude higher than in the work of \citeauthor{Nomura2009}, however their accretion rates are for the disc alone, while the rates here pertain to the embedded phase, which includes the envelope.  The discs presented in our work are inherently different from that of \citeauthor{Nomura2009}, since these discs are still forming. Therefore, the parcels tend to move quickly within the discs and material does not stay in the inner gas-phase rich zone long enough. Furthermore, \citeauthor{Nomura2009} used the variable Eddington factor method to calculate dust temperatures. It is not clear what the temperature differences are in comparison with this work, since only one stream line is published. In addition, \citeauthor{Nomura2009} and \citet{Walsh2014a} include viscous dissipation in the disc, which provides extra heating in the inner few AU.

More recently, \citet{Walsh2014b} modelled the abundances of complex organic molecules along multiple streamlines in a disc. The authors found that methanol was preserved along the midplane, prior to crossing of its snowline. At larger scale heights, a balance of photodesorption and photodissociation determines the fate of methanol molecules. Their findings are consistent with the conclusion reached in this work.

\subsection{Application to comets}
\label{comets}

A key application of the results obtained in this work is towards comets, especially in light of upcoming data from the \textit{Rosetta} mission. Comets likely form in young protoplanetary discs, and therefore may probe physical and chemical conditions in the early Solar Nebula. Efforts have been made to compare the abundances of various solid species relative to water ice in comets and towards young protostars. An example of such work is fig. $13$ in \citet{Oberg2011c2d}, using cometary data from \citet{MummaCharnley2011}, where cometary methanol abundances are shown to overlap with observations of methanol towards low-mass protostars. These simularities have also been demonstrated by \citet{Bockelee-Morvan2000}. It is not clear whether this is just a coincidence or if most of the cometary methanol is still prestellar from the cloud, out of which our Solar System formed. This point is also strongly dependent on how much methanol ice is built up during in the precollapse phase. \citet{Walsh2014a} concluded based on their fig.~$10$ that further disc formation is necessary to reproduce the observed cometary abundances of complex organic molecules, including methanol, as modelled cloud values are too low.

\citeauthor{Oberg2011c2d} concluded that the spreads in cometary and prestellar methanol ice abundances are comparable, with roughly a factor of $2$ difference in logarithmic space relative to water ice. Recalling Fig.~\ref{fgr:methbudget}, in the spread-dominated case, $10\lesssim R \lesssim30$ and $z\lesssim2$ is the zone with $\sim1-10^{-2}$ times less methanol ice than in the precollapse phase. In the infall-dominated case, a similar level of methanol ice loss is seen for the same $R$ and $z$ range. This implies that the factor of $2$ found by \citeauthor{Oberg2011c2d} is also found in these discs, and is a result of ice processing en route into the protoplanetary disc. Other zones in these discs allow for a loss greater than this factor of $2$, which is in agreement with methanol-poor cometary observations \citep{MummaCharnley2011}.

\section{Conclusions}
\label{conclusions}

This work employs an axisymmetric 2D semi-analytical physical model to simulate the formation of a low-mass protostar \citep{Visser2009,Visser2011}. Infall trajectories of parcels of matter are computed so that the material that makes up the protoplanetary disc can be sampled. A chemical model with a large gas-grain network is then applied to compute the chemical evolution of the system \citep{McElroy2013, Garrod2008, Walsh2014a}. First, the quiescent prestellar phase is simulated, and then the chemistry is computed along infall trajectories terminating at different points in the disc. By combining the two models into a physicochemical simulation, the chemical evolution as a function of physical evolution is studied for a star-forming system. Furthermore, two physical scenarios that vary in their respective dominant disc growth mechanism are studied side by side.

The main conclusion obtained in this work are as follows.
\begin{itemize}
\item The infall path sets the dust temperature and the UV flux, thus thermal desorption, photodesorption and photodissociation rates are trajectory-dependent. Methanol ice is stripped from the grains, if the dust temperature exceeds $\sim100$ K. If the UV flux is on the order of the interstellar field strength, methanol is rapidly destroyed both in the gas and solid phases. The abundance of methanol entering protoplanetary discs depends on how it is transported.
\item In the spread-dominated disc case, the inner $\sim5$ AU along the midplane is devoid of methanol ice and gas, since, en route, it is thermally desorbed, or photodissociated in the ice followed by thermal desorption of its photofragments, or photodissociated in the gas. Conversely, gas-phase methanol is plentiful for the infall-dominated disc case in the same region and methanol ice is lost solely due to thermal desorption.
\item The extended outer disc regions are rich in methanol ice that has been enhanced in the envelope.
\item The fate of the midplane depends on the mechanism, by which the disc grows. In the spread-dominated scenario, the midplane is methanol-ice poor and is built up of readsorbed prestellar methanol ice. In the infall-dominated disc case, the midplane is methanol-ice rich. It contains envelope-enriched and disc-formed methanol ice, which is unique to this colder disc, in addition to the preserved prestellar layers. 
\end{itemize}

The simulations show that the abundance and history of one of the key precursors for complex organic molecules, methanol, varies across protoplanetary discs, the physical structure of which is heavily influenced by the initial cloud rotation from which the star-disk system forms. The complex organics budget is directly affected by the availability of methanol, because it dissociates into radicals that are needed for synthesising larger, more complex species. The formation and survival of the latter will be studied in future work. The presence or absence of methanol and complex organic molecules determines the initial chemical composition of early protoplanetary and cometary material. The results obtained in this work are able to replicate the observed methanol cometary abundances and are also not inconsistent with the scenario of comets containing mostly prestellar methanol ice with some loss along the way.

\section{Acknowledgements}
\label{acknowledgements}
This work is supported by a Huygens fellowship from Leiden University, by the Netherlands Research School for Astronomy (NOVA), by a Royal Netherlands Academy of Arts and Sciences (KNAW) professor prize, by European Union A-ERC grant 291141 CHEMPLAN, and by the Netherlands Organization for Scientific Research (NWO, grant 639.041.335). R.V. is supported by NASA through an award issued by JPL/Caltech and by the National Science Foundation under grant 1008800.

\bibliographystyle{mn2e}
\bibliography{mybib} 
\bsp 

\label{lastpage}

\end{document}